\documentclass[amsmath,amssymb,pre,twocolumn,a4paper]{revtex4}
\usepackage{amsfonts} 
\usepackage{amsmath}
\usepackage{graphicx} 
\usepackage{overpic} 
\usepackage{color}
\usepackage[left=1.5cm, right=1.5cm, top=1.785cm, bottom=2.0cm]{geometry}
\usepackage{xr}
\newcommand{\be}{\begin{eqnarray}} 
	\newcommand{\ee}{\end{eqnarray}}

\begin{document}
	
	\title{Structural analysis of disordered dimer packings}
	\author{Esma Kurban and Adrian Baule\footnote{Correspondence to: a.baule@qmul.ac.uk}}
	\
	
	\affiliation{School of Mathematical Sciences, Queen Mary University of London, London E1 4NS, UK}

	\begin{abstract}
		Jammed disordered packings of non-spherical particles show significant variation in the packing density as a function of particle shape for a given packing protocol. Rotationally symmetric elongated shapes such as ellipsoids, spherocylinders, and dimers, e.g., pack significantly denser than spheres over a narrow range of aspect ratios, exhibiting a characteristic peak at aspect ratios of $\alpha_{\rm max}\approx 1.4-1.5$. However, the structural features that underlie this non-monotonic behaviour in the packing density are unknown. Here, we study disordered packings of frictionless dimers in three dimensions generated by a gravitational pouring protocol in LAMMPS. Focusing on the characteristics of contacts as well as orientational and translational order metrics, we identify a number of structural features that accompany the formation of maximally dense packings as the dimer aspect ratio $\alpha$ is varied from the spherical limit. Our results highlight that dimer packings undergo significant structural changes as $\alpha$ increases up to $\alpha_{\rm max}$ manifest in the reorganisation of the contact configurations between neighbouring dimers, increasing nematic order, and decreasing local translational order. Remarkably, for $\alpha>\alpha_{\rm max}$ our metrics remain largely unchanged, indicating that the peak in the packing density is related to the interplay of structural rearrangements for $\alpha<\alpha_{\rm max}$ and subsequent excluded volume effects with unchanged structure for $\alpha>\alpha_{\rm max}$.
	\end{abstract}	
	
	\maketitle

\section*{Introduction}
Jammed disordered particle packings have been used as a model to understand the structures of liquid crystals, glasses, self-assembly of nanoparticles, biological systems and granular media \cite{Torquato2010}. While there has been considerable recent progress in our understanding of jammed sphere packings \cite{Charbonneau:2017aa}, the effect of particle shape on the properties of jammed packings has been much less explored \cite{Baule:2014aa}. Considering one of the simplest macroscopic observables of packings --- the packing density --- one finds that many non-spherical shapes pack denser than spheres, which achieve maximal packing densities of $\phi_{\rm j}\approx 0.64$ for a wide range of packing protocols (although denser packings can also be achieved for specific protocols, see the discussion in \cite{Baule:2018aa}). For example, many polyhedra \cite{Haji-Akbari:2009aa,Jiao2011,Damasceno2012,Shepherd2012,Liu2017}, ellipsoids \cite{Buchalter1994,Delaney2011,Donev2004,Man2005}, spherocylinders \cite{Williams2003,Zhao:2012aa,Abreu:2003aa,Jia:2007aa,Bargiel:2008aa,Wouterse:2009aa,Kyrylyuk:2011aa}, and dimers \cite{Faure2009,Shiraishi2020}, as well as irregular shapes such as those composed of a number of overlapping spheres \cite{Miskin2014,Roth2016} achieve packing densities $\phi_{\rm j}\ge 0.7$, with the densest disordered packings so far found for tetrahedra at $\phi_{\rm j}\approx 0.78$ \cite{Haji-Akbari:2009aa}. Plotting the packing density as a function of a continuous shape descriptor, such as the aspect ratio $\alpha$ (for rotationally symmetric elongated shapes), exhibits a non-monotonic behaviour with a peak at $\alpha\approx1.4-1.5$ for ellipsoids, spherocylinders, and dimers, with some variations due to the packing protocol. For larger aspect ratios, the packing density decreases, following, e.g., an approximate scaling behaviour $\phi_{\rm j}\sim1/\alpha$ for spherocylinders \cite{Philipse:1996aa}.  

In this study, we revisit dimer packings simulated with the MD platform LAMMPS using a gravitational pouring protocol. Our goal is to identify structural features that characterize the peak in the packing density by focusing on details of the contact statistics as well as positional and orientational order metrics. In this context, it is important to emphasize the role of the protocol in the packing generation. The interplay between the packing density and the degree of order that arises by tuning the protocol parameters has been widely discussed for spheres, most notably in the critique of the well-posedness of the concept of ``random close packings" \cite{Torquato:2000aa}. For non-spherical particles, the protocol dependence is manifest in the relatively large variance of results reported for $\phi_{\rm j}$ for the same shape, e.g., for spherocylinders \cite{Williams2003,Zhao:2012aa,Abreu:2003aa,Jia:2007aa,Bargiel:2008aa,Wouterse:2009aa,Kyrylyuk:2011aa}. Our viewpoint is thus to focus on packings generated by a specific protocol, namely the widely used pouring under gravity, and understand how shape variation changes the structural features of these packings.

Previous studies of ordering effects in random packings of elongated particles obtained inconsistent results, which might be due to different protocols and boundary conditions used. For example, simulations of prolate ellipsoids by pouring into a container under gravity found considerable nematic order, whereby the ellipsoids' symmetry axes (the semi-major axes) tend to lie within the plane normal to the gravity direction \cite{Buchalter1994,Delaney2011,Gan2020}. This ordering effect has been explained as a result of the particles' tendency to minimize the gravitational potential energy \cite{Buchalter1994}. On the other hand, simulations that compress or inflate the non-spherical particles from an initial random state such as the Lubachevsky-Stillinger algorithm (applied to ellipsoids \cite{Donev2004,Donev2007}) or a mechanical contraction algorithm (applied to spherocylinders \cite{Wouterse2007,Wouterse2009,Cordova2014}) do not find any significant order as is also observed with other geometric simulation methods \cite{Zhao2012,Meng2016}. While 3D experiments of ellipsoids \cite{Man2005} and elongated colloids \cite{Sacanna2007} did not observe any signatures of order, experiments of asymmetric dumbbells in 2D showed strong orientational correlations between neighbours due to mutual restrictions on positions \cite{Han2012}. The order characteristics of dimers in 3D have so far not been investigated to our knowledge.

The remainder of this article is organized as follows: in Section~\ref{Sec:simulation} we present the details of our simulation method with LAMMPS. In Section.~\ref{Sec:results}, we present results on our analysis of the packing fraction, contact number, and orientational/positional order metrics. Finally, we conclude in Section~\ref{Sec:conclusion} with a discussion of our results.

\section{Simulation Method}
\label{Sec:simulation}

\begin{figure}
	\centering	
	(a)\includegraphics[height=1.5cm]{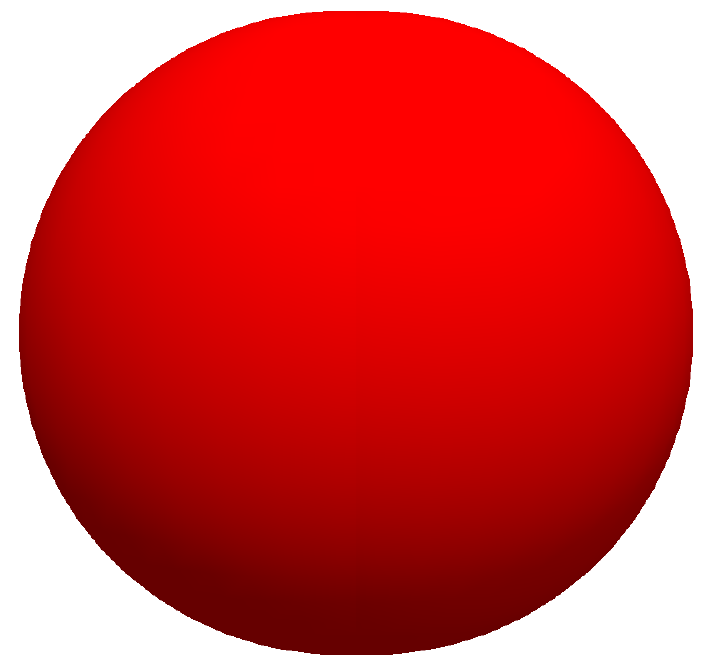} 
	(b)\includegraphics[height=1.5cm]{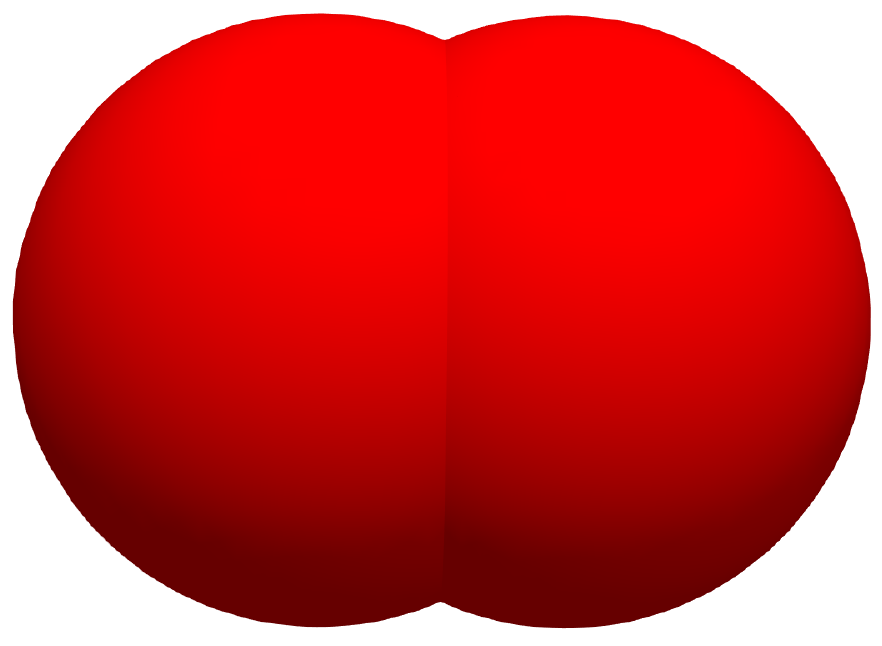} 
	(c)\includegraphics[height=1.5cm]{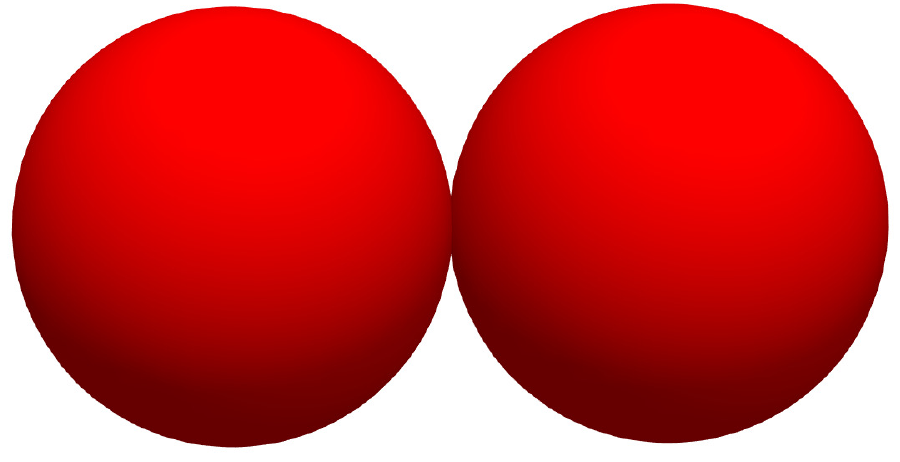}
	\caption{\label{Fig:dimers}Dimer shape defined by the aspect ratio: (a) $\alpha=1.05$, (b) $\alpha=1.4$, (c) $\alpha=2$.}
\end{figure}

Disordered packings of monodisperse frictionless symmetric dimers in three dimensions are generated with the molecular dynamics platform LAMMPS \cite{Plimpton1995,LAMMPS}. The dimers are obtained by overlapping two identical spheres with diameter $d$ and mass $m$. We study dimers with aspect ratios $\alpha$ in the range $1.0005\leq\alpha\leq2$, where $\alpha$ is given as the ratio of the length over the width, see Fig.~\ref{Fig:dimers}. In this packing protocol, $N=12,000-15,000$ monodisperse dimers are poured under gravity into a three-dimensional box of side length $\approx 20d$. The lateral ($\hat{\mathbf{x}}$-$\hat{\mathbf{y}}$-plane) boundary conditions are chosen to be periodic and the box is bounded in the $\hat{\mathbf{z}}$-direction by a rough surface at the bottom (implemented by the ``fix wall/gran hertz/history" command). During a simulation run, a gravitational force acts on the dimers in the $\hat{\mathbf{z}}$-direction. The pouring protocol makes use of LAMMPS' ``fix pour" command, which repeatedly inserts particles into the simulation box every few timesteps within a specified insertion region $30-40d$ above the bottom and releases them until $N$ particles have been added overall. In the insertion region, particles are added with random positions and orientations and without any overlap. Particles are only inserted again after the previously inserted particles have fallen out of the insertion region under the gravitational force.

LAMMPS treats a dimer defined by a fixed distance between its two constituent spheres as an independent rigid body (implemented by the ``fix rigid/small" command). The total force and torque on each dimer rigid body are computed as the sum of the forces and torques on its constituent spheres in every time step. The coordinates, velocities, and orientations of the constituent spheres are then updated so that the dimer moves and rotates as a single entity.

LAMMPS can natively implement different models for calculating the contact forces between the spheres. In this study, a Hookean model is chosen because of its convenience to dissipate residual kinetic energy and hence to reach a static state quickly \cite{Silbert2001}. In the Hookean model, when two spheres $i$ and $j$ having positions $\mathbf{r}_i$ and $\mathbf{r}_j$, respectively, are in contact, they experience a relative normal compression with overlap $\delta=d-r_{ij}$, where $\mathbf{r}_{ij}=\mathbf{r}_i-\mathbf{r}_j$ and $r_{ij}=|\mathbf{r}_{ij}|$. The resulting force is $\mathbf{F}_{ij}=\mathbf{F}_{ij}^{\rm{n}}+\mathbf{F}_{ij}^{\rm{t}}$, where $\mathbf{F}_{ij}^{\rm{n},\rm{t}}$ are the normal and tangential contact forces, respectively, given as \cite{Silbert2001}:
\begin{equation}
	\mathbf{F}_{ij}^{\rm{n}}=K_{\rm n}\delta\, \mathbf{n}_{ij}-\frac{m}{2}\gamma_{\rm n} \mathbf{v}_{\rm n} \qquad \qquad
	\mathbf{F}_{ij}^{\rm{t}}=-K_{\rm t}\Delta\mathbf{s}_{t}-\frac{m}{2}\gamma_{\rm t} \mathbf{v}_{\rm t}.
\end{equation}
Here, $\mathbf{n}_{ij}=\mathbf{r}_{ij}/r_{ij}$, $\mathbf{v}_{\rm n,t}$ are the normal and the tangential components of the relative velocity of the spheres $i$ and $j$, and $K_{\rm n,t}$ and $\gamma_{\rm n,t}$ are the elastic and viscoelastic constants, respectively. The quantity $\varDelta \mathbf{s}_{\rm t}$ denotes the elastic tangential displacement between the spheres \cite{Silbert2001}. The total force $\mathbf{F}_{i}^{\rm{tot}}$ on sphere $i$ in a gravitational field $\mathbf{g}=-g\,\hat{\mathbf{z}}$ is then given as:
	\begin{equation}
		\mathbf{F}_{i}^{\rm{tot}}=m\,\mathbf{g}+\sum_{i\neq j}\mathbf{F}_{ij}^{\rm{n}}+\sum_{i\neq j}\mathbf{F}_{ij}^{\rm{t}},
	\end{equation}
where the sum runs over all $j$ spheres in contact with sphere $i$.

Throughout the investigation we set our basic units as $d=1$, $m=\pi/6$, and $g=1$. Distances, times, velocities, forces and elastic constants are then measured in units of $d$, $\sqrt{d/g}$, $\sqrt{gd}$, $mg$, $mg/d$, respectively. We generally use $K_{\rm n}=2\times 10^5 mg/d$ unless otherwise indicated. Additionally, we simulate also harder dimers with a normal spring constant $K_{\rm n}=2\times 10^6 mg/d$ and softer ones with $K_{\rm n}=2\times 10^4 mg/d$ to examine the effect of particle hardness on the contact number of the dimers at small aspect ratios. We set $\gamma_{\rm t}=0$ and the remaining parameters used are given in Table~\ref{table}. The choice of most of these values follows the discussion in \cite{Silbert2002}.

\begin{table}
\small
    \caption{Material parameter values and time step $\Delta t$ used in the simulations}
    \label{table}
	\centering
	\begin{tabular*}{0.48\textwidth}{@{\extracolsep{\fill}}c|c|c|c}
		$K_{\rm n}$ $(mg/d)$ & $K_{\rm t}/K_{\rm n}$ & $\gamma_{\rm n}$ $(mg/d)$ & $\Delta t$ $(\sqrt{d/g})$ \\
		\hline
		$2\times 10^4$ & 2/7 & 15  & 0.003 \\
		$2\times 10^5$ & 2/7 & 50 & 0.001 \\
		$2\times 10^6$ & 2/7 & 150 & 0.0003 \\
	\end{tabular*}	
\end{table}

Simulations are run until the system reaches a static equilibrium when the kinetic energy per particle is less than $10^{-8} mgd$ for small $K_{\rm n}$ and up to three orders of magnitude less for large $K_{\rm n}$. For example, when $K_{\rm n}=2\times 10^5 mg/d$ the simulation takes $3-8 \times 10^6\Delta t$ to reach equilibrium, which depends on the chosen aspect ratio and also on the random initial configurations when particles are poured into the container. For further details of any of the LAMMPS commands used, we refer the reader to the LAMMPS documentation \cite{LAMMPS}.

\section{Structural Analysis}

\label{Sec:results}

\subsection{Packing Fraction}

\begin{figure}
	\centering
	\includegraphics[width=0.3\textwidth]{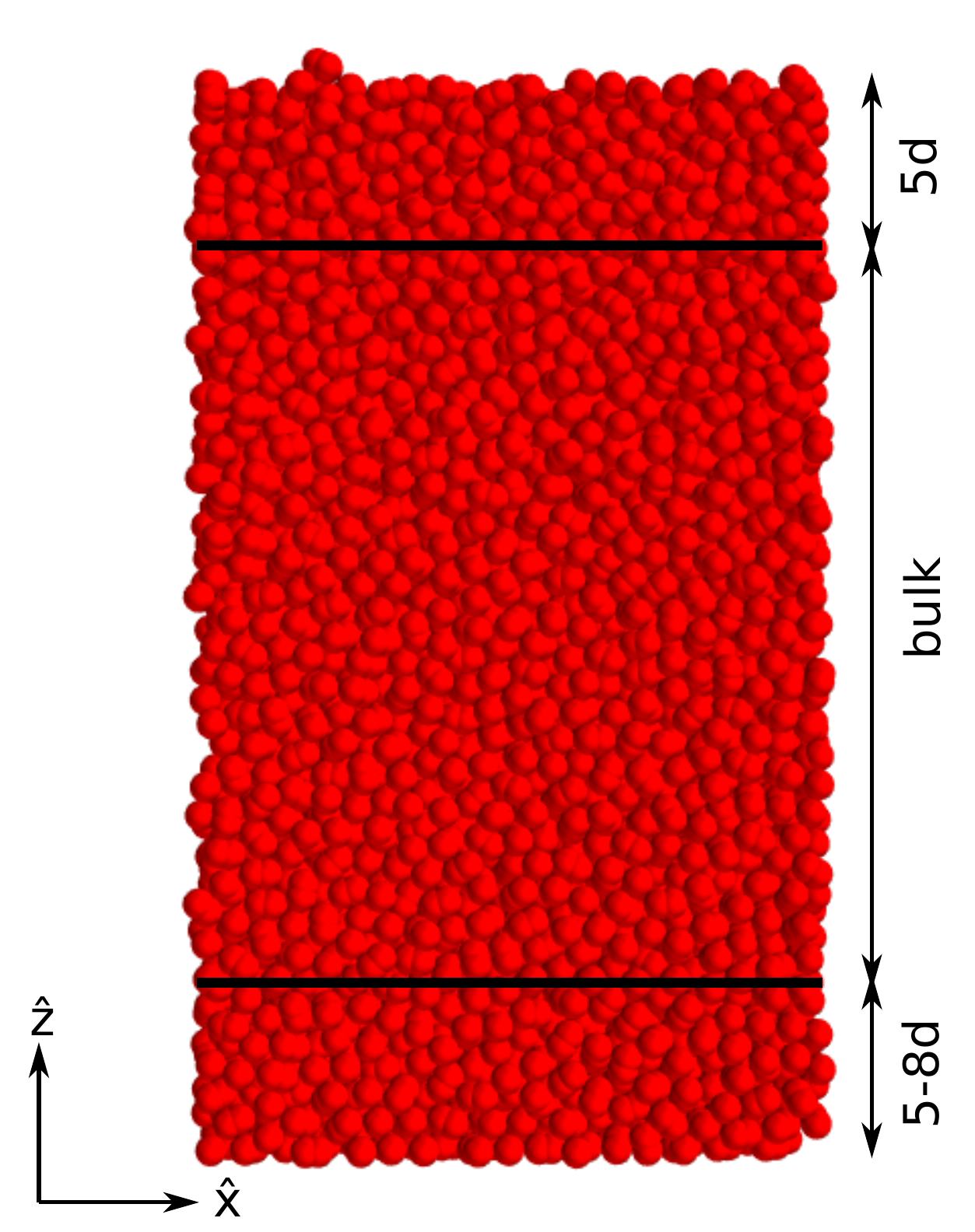}
	\caption{\label{Fig:bulk}The bulk region shown in the $\hat{\mathbf{x}}$-$\hat{\mathbf{z}}$-plane.}
\end{figure}

We calculate the packing fraction of the dimer packings for various aspect ratios. The packing density is determined for the bulk region shown in Fig.~\ref{Fig:bulk}. The particles within $5-8d$ from the container floor are excluded from the bulk region since they can be highly crystallized. The thickness of this crystallized region depends on many factors such as the box dimension and the pouring height. Excluding the particles within $5-8d$ provides results that are largely unaffected by the crystallization. The particles within $5d$ from the upper-most particles have also been excluded from the bulk because their Voronoi volumes can not be decided accurately due to deficiencies in their neighbourhood.

In order to determine the packing density in the bulk region, we calculate the Voronoi volume of each dimer in the bulk, which is defined as the space that is closer to the surface of a given dimer than to that of any other dimer. While a formal parametrization of the Voronoi volume of a dimer is analytically tractable \cite{Baule2013}, a straightforward computational method makes use of LAMMPS' built-in routine to determine the Voronoi volume of the individual spheres in the packing using a conventional Voronoi tessellation. The Voronoi volume $W_i$ of a dimer is then found by summing the Voronoi volumes of its two constituent spheres. The bulk volume $V_{\rm b}$ occupied by $N_{\rm b}$ dimers in the bulk is calculated as $V_{\rm b}=\sum_{i=1}^{N_{\rm b}} W_i$. We then obtain the packing fraction as $\phi_{\rm j}=N_{\rm b}V_{\alpha}/V_{\rm b}$ where $V_{\alpha}$ is the volume of a dimer with aspect ratio $\alpha$. The volume $V_\alpha$ is found by subtracting the overlap volume from the sum of its constituent sphere volumes. The overlap volume contains two equal spherical caps whose volume can be calculated exactly, see Appendix~\ref{App:overlap}. Note that a dimer is considered to be part of the bulk region only if the centres of both constituent spheres are within the bulk. All average quantities discussed in the following are calculated for dimers in the bulk only.

\begin{figure}
	\centering
	\includegraphics[width=0.45\textwidth]{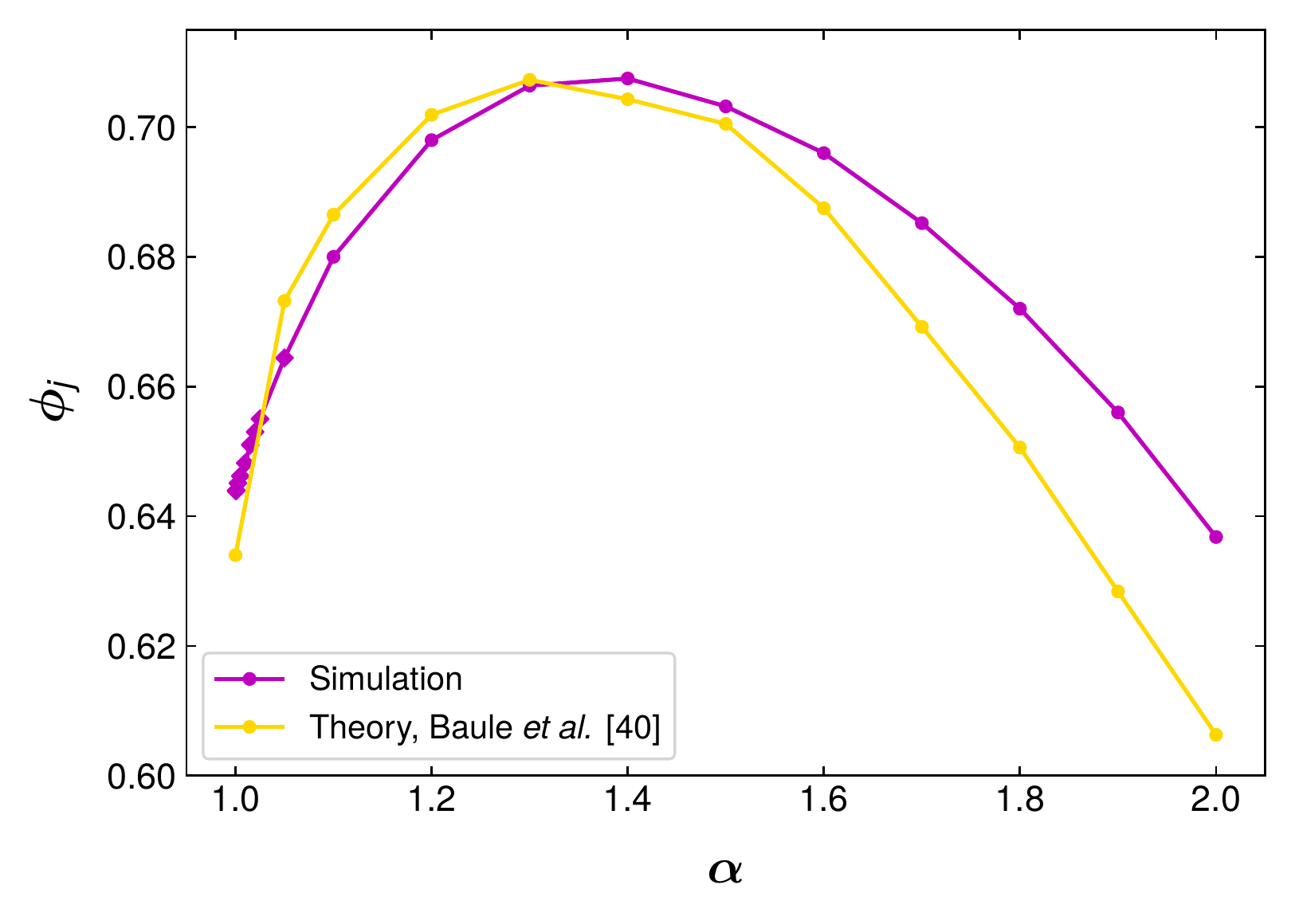}
	\caption{\label{Fig:phi}The packing fraction $\phi_{\rm j}$ as a function of the dimer aspect ratio $\alpha$. Simulation values of $\phi_{\rm j}$ are shown averaged over 10 independent simulation runs for $\alpha\ge 1.1$ (dots), and for a single run for $\alpha<1.1$ (diamonds).}
\end{figure}

We plot the packing fraction $\phi_{\rm j}$ of the dimers as a function of the aspect ratio $\alpha$ in Fig.~\ref{Fig:phi}. As can be seen from Fig.~\ref{Fig:phi}, the packing fraction $\phi_{\rm j}$ has a non-monotonic relationship with $\alpha$, i.e., it increases as $\alpha$ increases until reaching a peak at $\phi_{\rm j}=0.707$ for $\alpha=\alpha_{\rm max}=1.4$, beyond that it decreases. These results are in agreement with previous studies \cite{Faure2009,Shiraishi2020} and also show reasonably good agreement with results from a mean-field calculation \cite{Baule2013}, shown in Fig.~\ref{Fig:phi}. Systematic deviations between our simulations and the mean field theory are in particular visible in the behaviour for larger aspect ratios $\alpha>1.5$, which are likely due to the strong mean-field assumptions. In fact, the mean-field theory relies on a reduction of higher-order positional correlations to pair correlations and also neglects orientational correlations between particles. The latter become more significant for particles of larger aspect ratios, see Sec.~\ref{Sec:S}.

\subsection{Contact and coordination numbers}

We introduce the {\it contact number} $z$ as the average number of contact points of a dimer and the {\it coordination number} $z_{\rm c}$ as the average number of neighbours of a dimer, whereby a neighbour is defined as another dimer with which at least one contact point is shared. While $z=z_{\rm c}$ for smooth convex shapes like spheres, ellipsoids, and spherocylinders, $z\ge z_{\rm c}$ for concave shapes like dimers, since two particles can share more than one contact point. In general, two dimers A and B share a contact point if the separation vector of two spheres $i$ and $j$, with sphere $i$ in dimer A and sphere $j$ in dimer B, satisfies $r_{ij}\le d$, which can be detected with high numerical precision. Two dimers can thus share up to four different contact points. Due to the soft interaction potential the contact ``point" is strictly a small overlap region, which creates some complications at small dimer aspect ratios, see below.

In Fig.~\ref{Fig:zvsalpha}(a) we show the behaviour of $z_{\rm c}$ as a function of $\alpha$ and the associated distributions of $z_{\rm c}$ for a set of aspect ratios. We observe a smooth increase of $z_{\rm c}(\alpha)$ for $\alpha>1$ with a maximum at $z_{\rm c}=8.34$ for $\alpha=1.5$ followed by a slight decay. The qualitative behaviour is in line with the results of \cite{Shiraishi2020}, where dimer packings were generated using an energy minimization protocol, although our values of $z_{\rm c}$ are consistently larger over the range of aspect ratios. The distributions $P(z_{\rm c})$ are approximately symmetric and Gaussian (Fig.~\ref{Fig:zvsalpha}(a,inset)).

\begin{figure*}
	\centering
	(a)\includegraphics[width=0.45\textwidth]{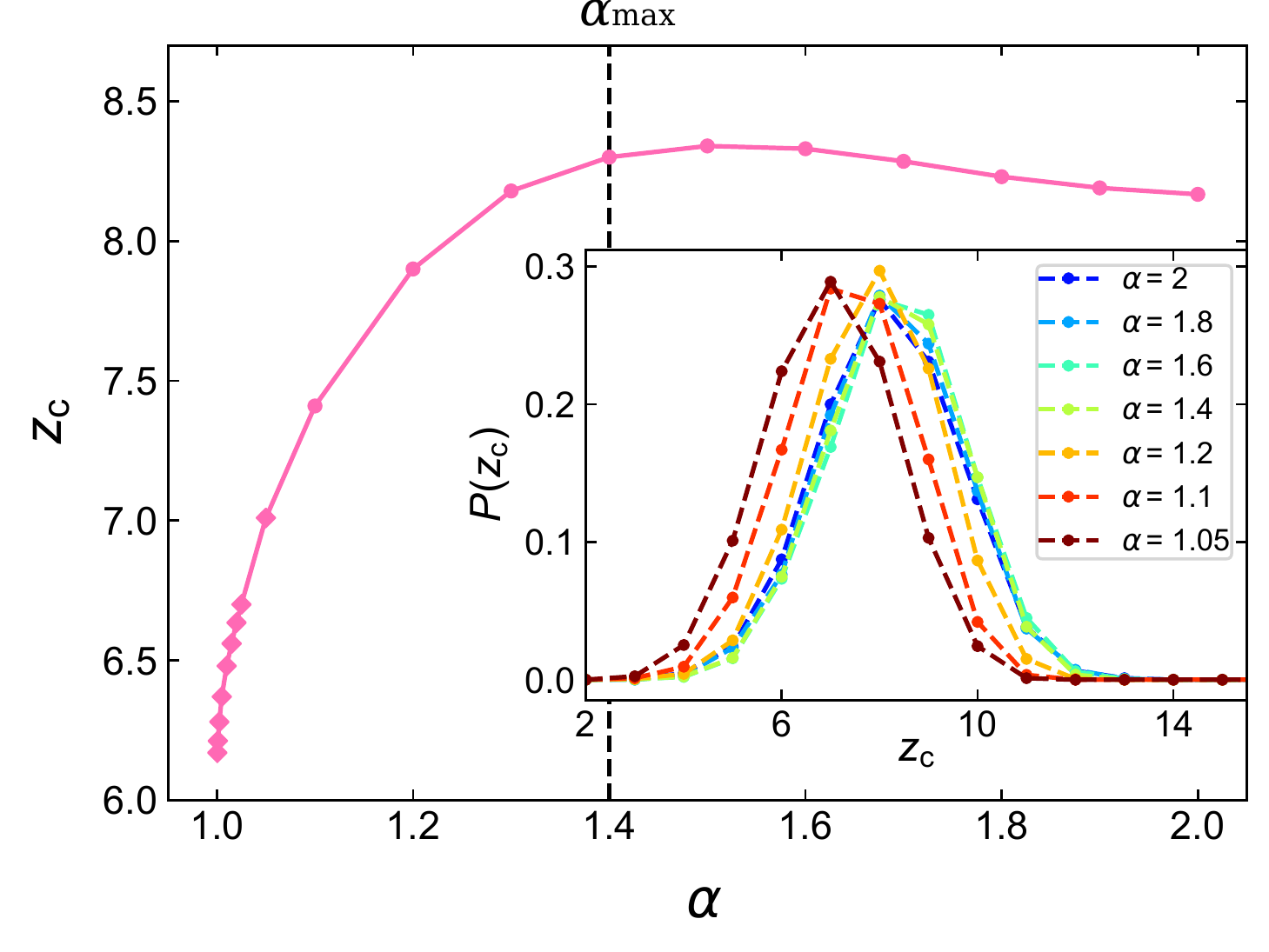}\hspace{0.5cm} (b)\includegraphics[width=0.45\textwidth]{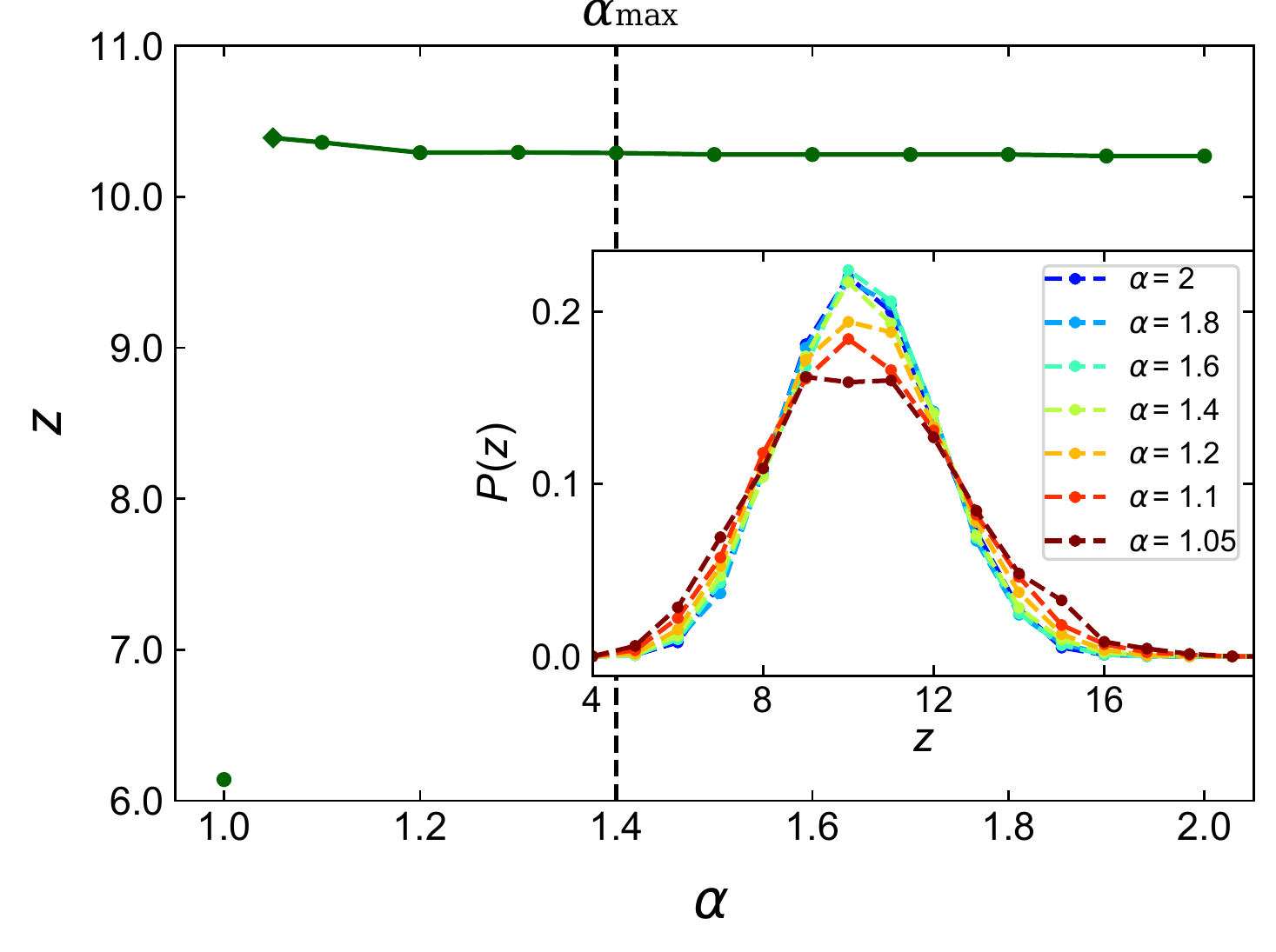} 
	\caption{\label{Fig:zvsalpha}(a) The coordination number $z_{\rm c}$ vs $\alpha$ and distributions $P(z_{\rm c})$ for various aspect ratios (inset). (b) The contact number $z$ vs $\alpha$ and distributions $P(z)$ (inset). The values of $z_{\rm c}$ and $z$ are shown averaged over 10 independent simulation runs for $\alpha\ge 1.1$ and $\alpha=1$ (dots), and for a single run for $1<\alpha<1.1$ (diamonds).}
\end{figure*}

On the other hand, the contact number $z$ does not exhibit such a smooth increase, see Fig.~\ref{Fig:zvsalpha}(b). First establishing the baseline for sphere packings at $\alpha=1$ with our protocol, we find that $z=6.14$ for spheres. This value is slightly above the isostatic value of $z=2d_{\rm f}=6$, where $d_{\rm f}$ denotes the degrees of freedom of a particle, generally found for disordered sphere packings using a variety of packing protocols \cite{Baule:2018aa}. We suspect that this difference is due to the gravitational packing protocol and the interaction potential with non-zero softness, see also the comparable values found in the studies of sphere packings \cite{Silbert2002,Faure2009} using a similar protocol. Deforming spheres into dimers, the smallest aspect ratio of dimers for which we are able to report the contact number reliably is $\alpha=1.05$, for which we find $z=10.39$. For larger aspect ratios, $z$ decreases slightly, but then remains unchanged at $z=10.28$ for $\alpha>1.2$. The difference with the isostatic value $z=2d_{\rm f}=10$ is approximately of the same magnitude as the difference for spheres using our packing protocol. By comparison, the studies in \cite{Schreck2010,Shiraishi2019,Shiraishi2020} find that dimers are almost exactly isostatic, which is thus in line with our findings. The observation of a constant $z$ for all aspect ratios of dimers is an important difference with the behaviour of convex elongated shapes such as ellipsoids and spherocylinders, which are hypostatic ($z<2d_f$) at small aspect ratios and show a smooth increase upon shape deformation from the sphere like the coordination number $z_{\rm c}$ here.

\begin{figure}[h]
	\centering
	(a)\includegraphics[height=4cm]{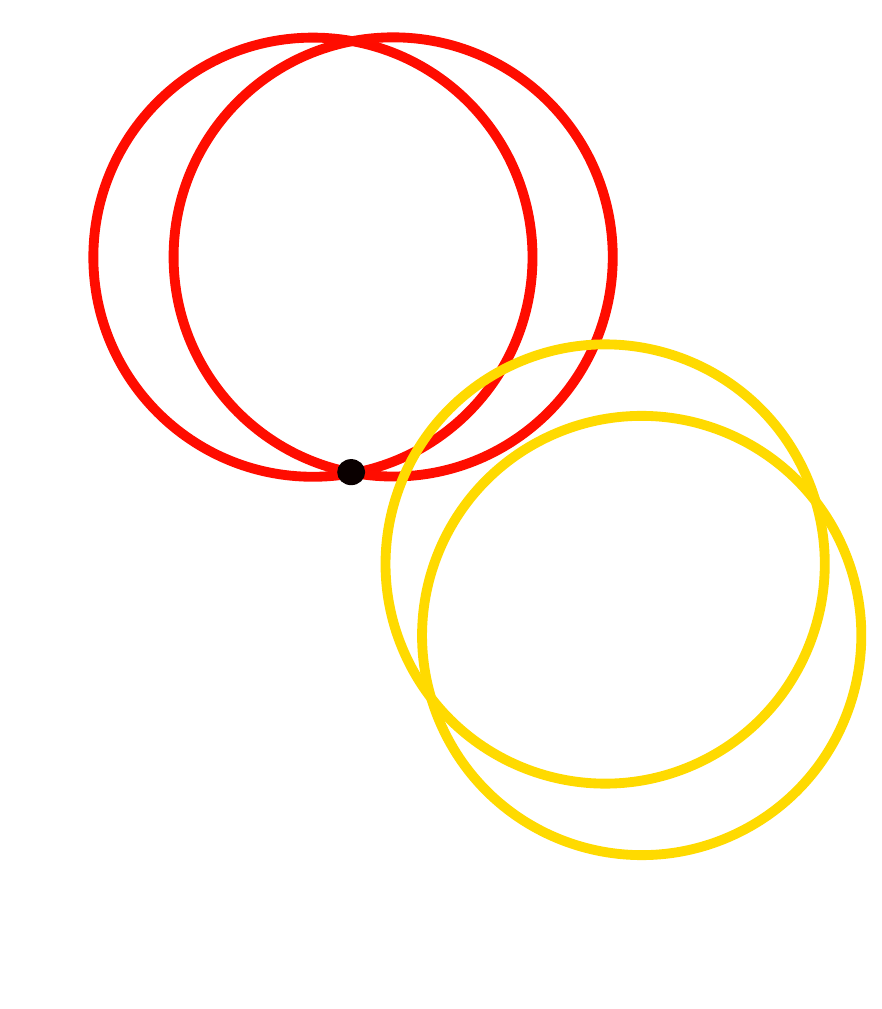}\hspace{0.5cm}
	(b)\includegraphics[height=4cm]{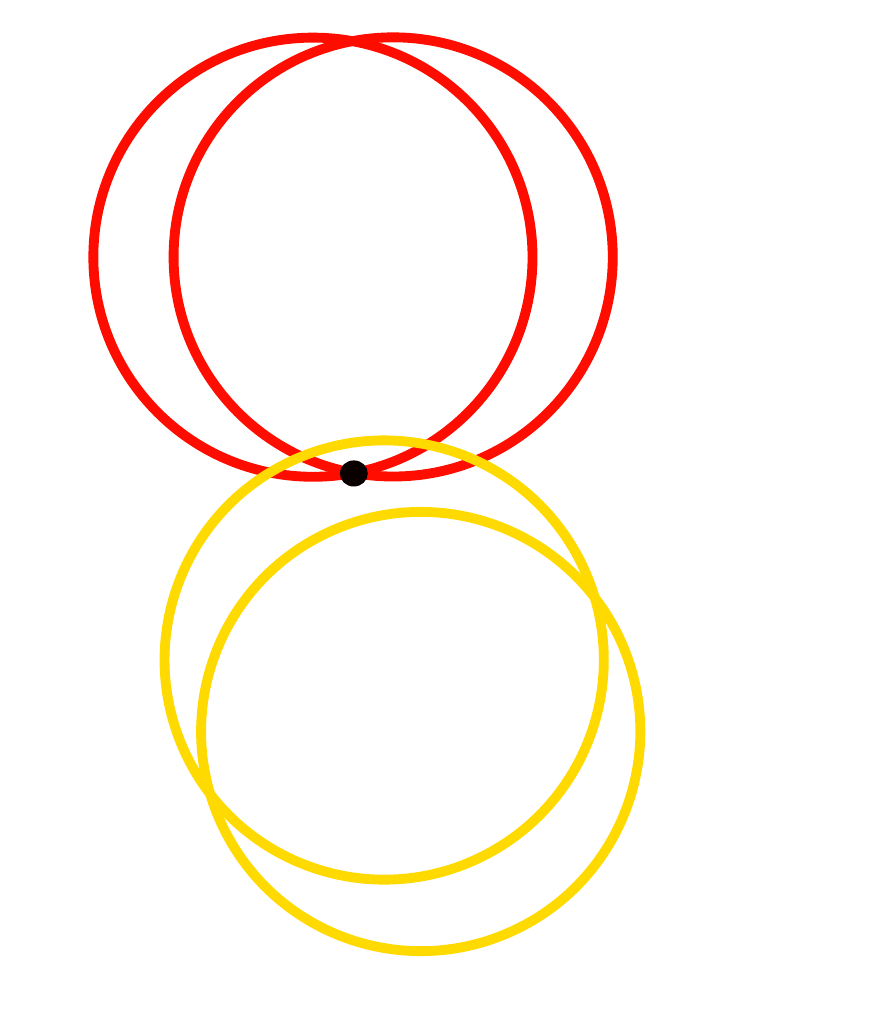}
	\caption{\label{Fig:DCcontact}Illustrations of ``double" and ``cusp" contacts shown in 2D as discussed in \cite{Shiraishi2019}. (a) Double contact: the yellow sphere is embedded into the red dimer so deeply that it contacts both red spheres. (b) Cusp contact: the yellow sphere contacts both red spheres by covering the cusp point (black point) of the red dimer.}
\end{figure}

We highlight that for very small aspect ratios $\alpha\in (1,1.05)$ the calculation of $z$ is unreliable, since our particle model leads to incorrect contact detections: the overlap regions due to the particle softness can extend far enough into the dimer as to create a contact with an interior sphere as illustrated in Fig.~\ref{Fig:DCcontact}.

\begin{figure}[h]
	\centering
	\includegraphics[width=0.45\textwidth]{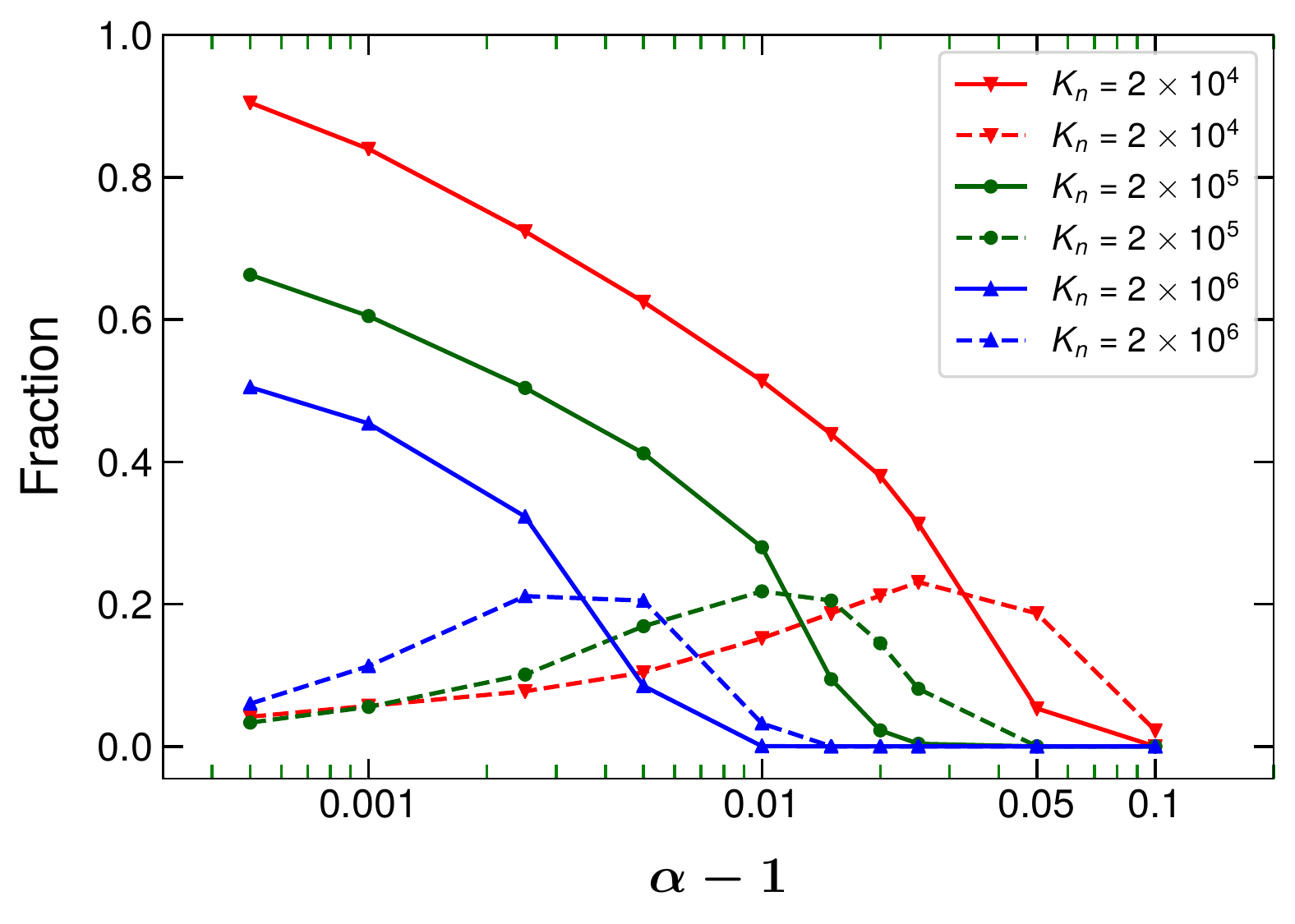}
	\caption{\label{Fig:Kneffect}The fraction of double (solid lines) and cusp contacts (dashed lines) in the dimer packings for small $\alpha$ and three normal spring constants $K_{\rm n}$.}
\end{figure}

Such problematic contact configurations for dimers were also identified in the recent work by Shiraishi {\it et al.} \cite{Shiraishi2019,Shiraishi2020} and separated into ``double" and ``cusp" contacts, see Fig.~\ref{Fig:DCcontact}. Shiraishi {\it et al.} investigated the contact number of dimer packings using a compression protocol with soft particle interactions for various packing densities $\phi$. For large enough values of the excess packing density $\Delta\phi=\phi-\phi_{\rm j}$, where $\phi_{\rm j}$ denotes the packing density at jamming onset, ``double" and ``cusp" contacts were observed. In their analysis, these contacts could thus be avoided by setting an upper limit for $\Delta\phi$ at each aspect ratio studied and they observed that this upper limit approaches zero as $\alpha\to 1$. In our case, the occurrence of these configurations depends on the stiffness value $K_{\rm n}$ as shown in Fig.~\ref{Fig:Kneffect}, where it can be seen that the threshold aspect ratio, at which double and cusp contacts occur, is shifted to smaller aspect ratios for larger $K_{\rm n}$. For any value of $K_{\rm n}$, double and cusp contacts will occur at sufficiently small aspect ratios and thus the contact number very close to the sphere shape can not be reliably established. For $K_{\rm n}=2\times10^5$ we see that double and cusp contacts do not occur for $\alpha\ge 1.05$, which is the lower limit of $\alpha$ used in our contact number analysis.

\begin{table*}
	\centering
	\caption{Five distinct contact configurations of two dimers. We show illustrations for aspect ratios $\alpha=1.2$ and $\alpha=2$. The total number of contact points for each type is: one (Type 1), two (Type 2,3), three (Type 4), four (Type 5)}
	\label{table2}
	\begin{tabular*}{\textwidth}{@{\extracolsep{\fill}}cccccc}
		\hline
		$\alpha$  & Type 1 & Type 2 & Type 3 & Type 4 & Type 5 \\
		\hline \\
		$1.2$  &	
		\begin{minipage}{.15\textwidth}
			\includegraphics[height=2.5cm]{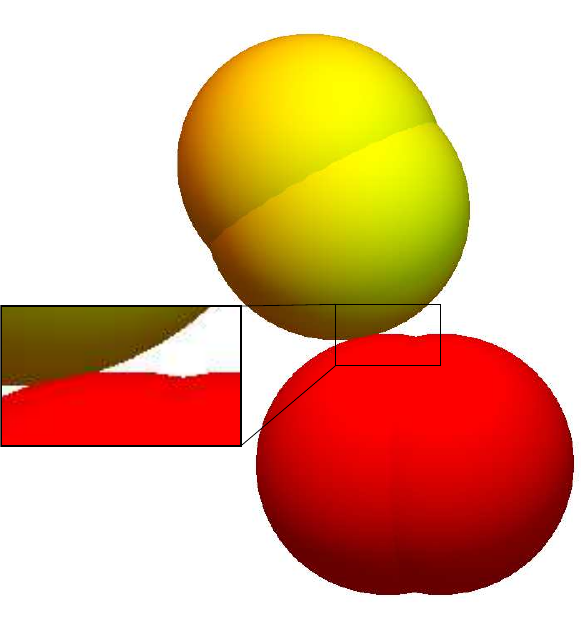}
		\end{minipage} 	
		&  
		\begin{minipage}{.15\textwidth}
			\includegraphics[height=2.5cm]{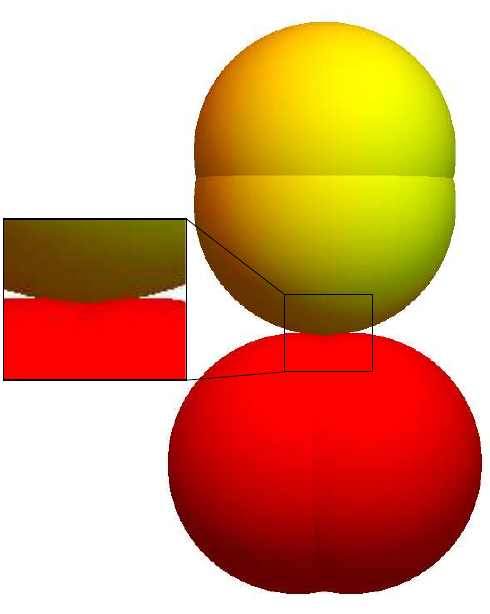} 
		\end{minipage}
		&
		\begin{minipage}{.15\textwidth}
			\includegraphics[height=2.5cm]{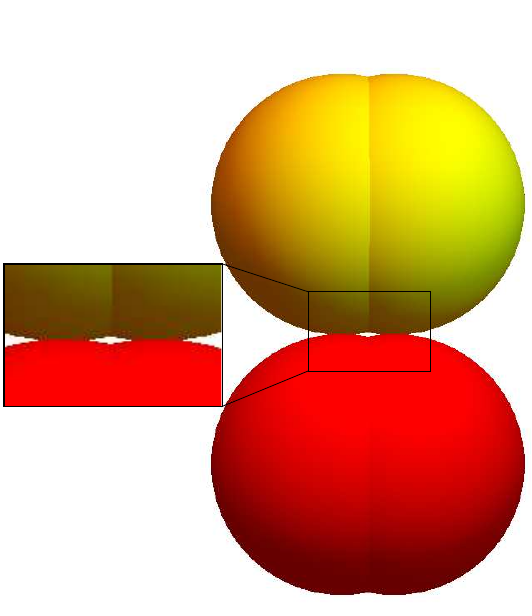} 
		\end{minipage}
		&   
		\begin{minipage}{.15\textwidth}
			\includegraphics[height=2.5cm]{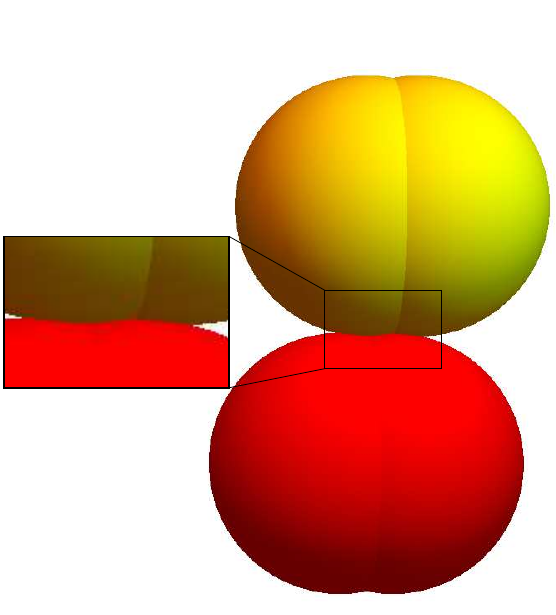} 
		\end{minipage}
		&
		\begin{minipage}{.15\textwidth}
			\includegraphics[height=2.5cm]{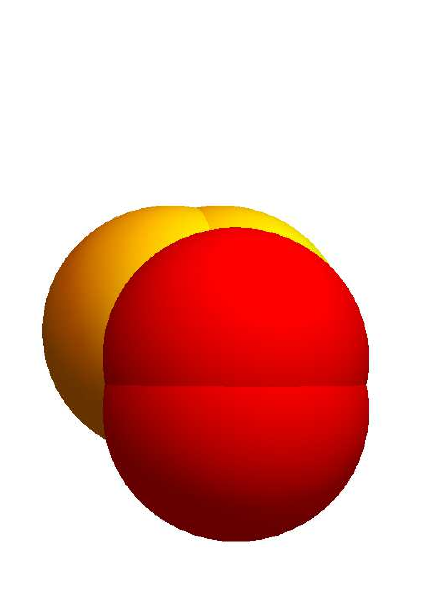} 
		\end{minipage}    
	     \\ 
		\hline\\
		$2$  &	
		\begin{minipage}{.15\textwidth}
			\includegraphics[height=3.5cm]{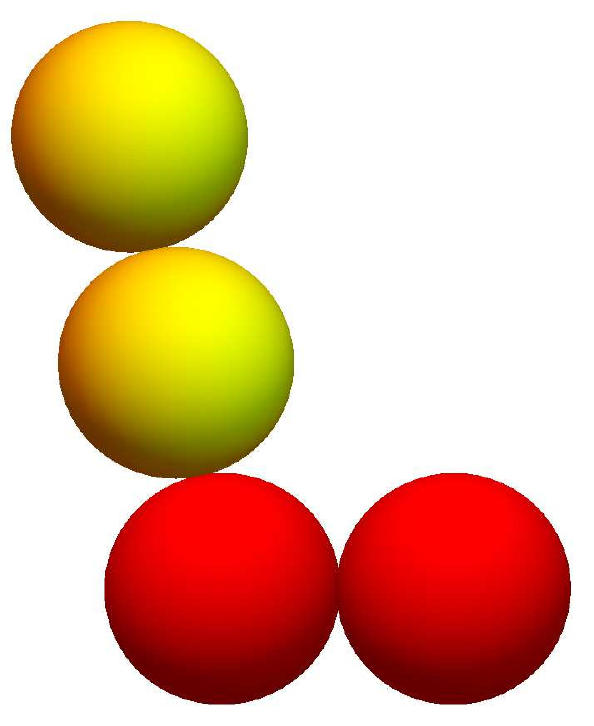}
		\end{minipage} 	
		&  
		\begin{minipage}{.15\textwidth}
			\includegraphics[height=3.5cm]{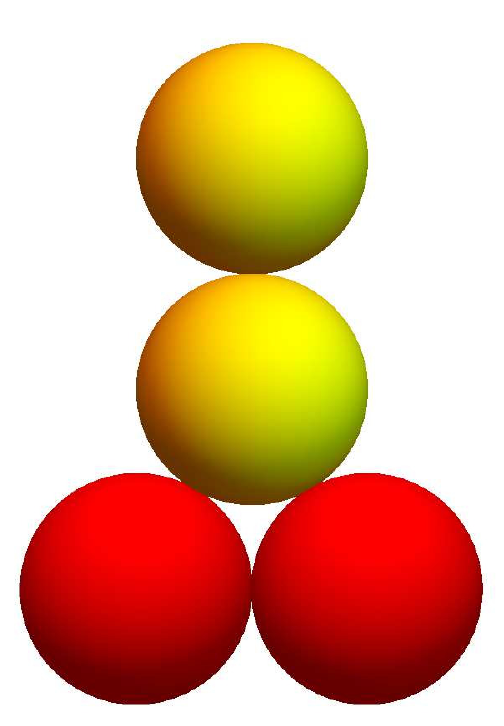} 
		\end{minipage}
		&
		\begin{minipage}{.15\textwidth}
			\includegraphics[height=3.5cm]{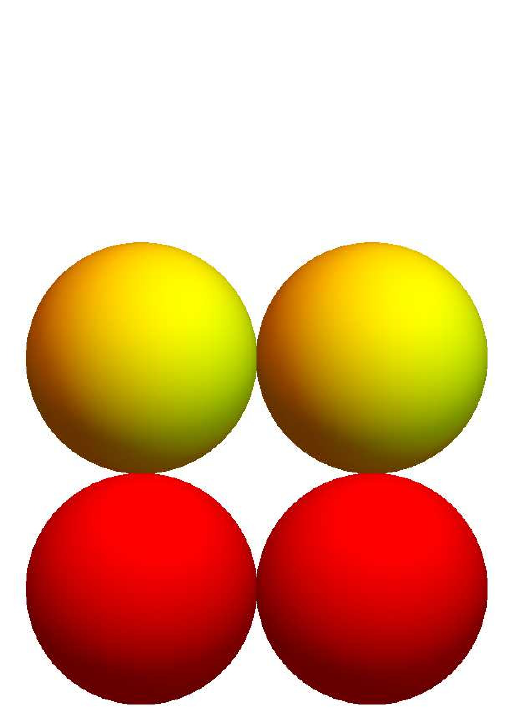} 
		\end{minipage}
		&   
		\begin{minipage}{.15\textwidth}
			\includegraphics[height=3.5cm]{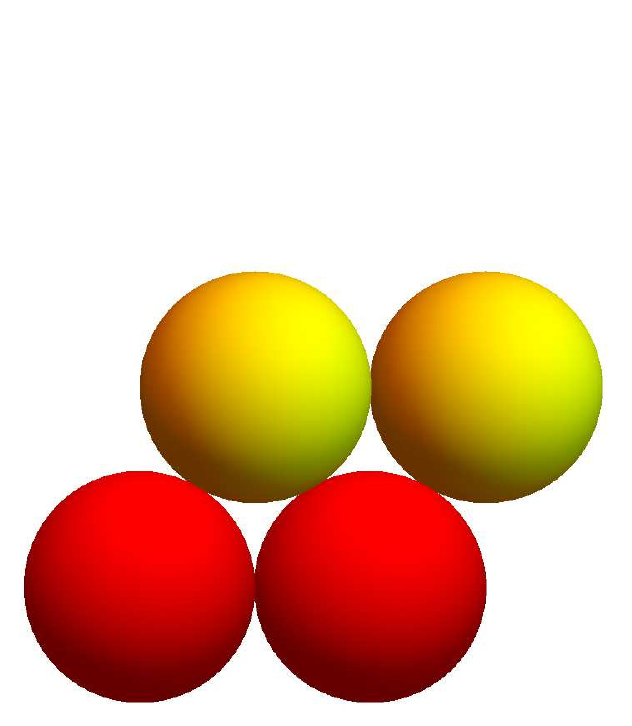} 
		\end{minipage}
		&
		\begin{minipage}{.15\textwidth}
			\includegraphics[height=3.5cm]{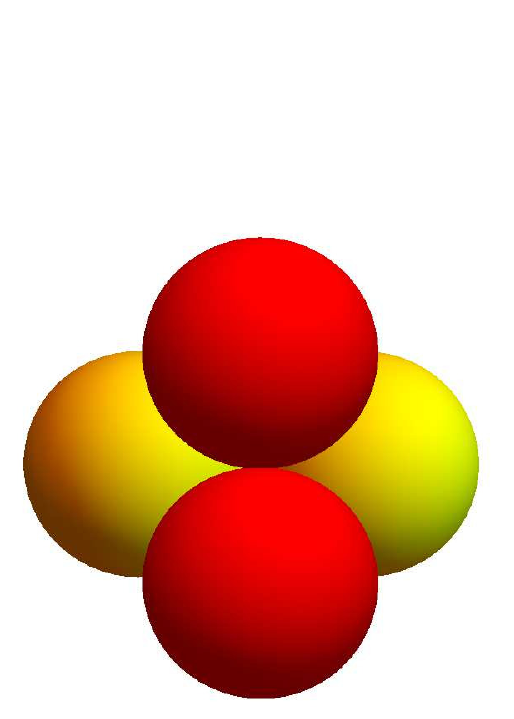} 
		\end{minipage}    
		\\ 
		\hline		
	\end{tabular*}	
\end{table*}

\begin{figure}
	\centering	
	\includegraphics[width=0.45\textwidth]{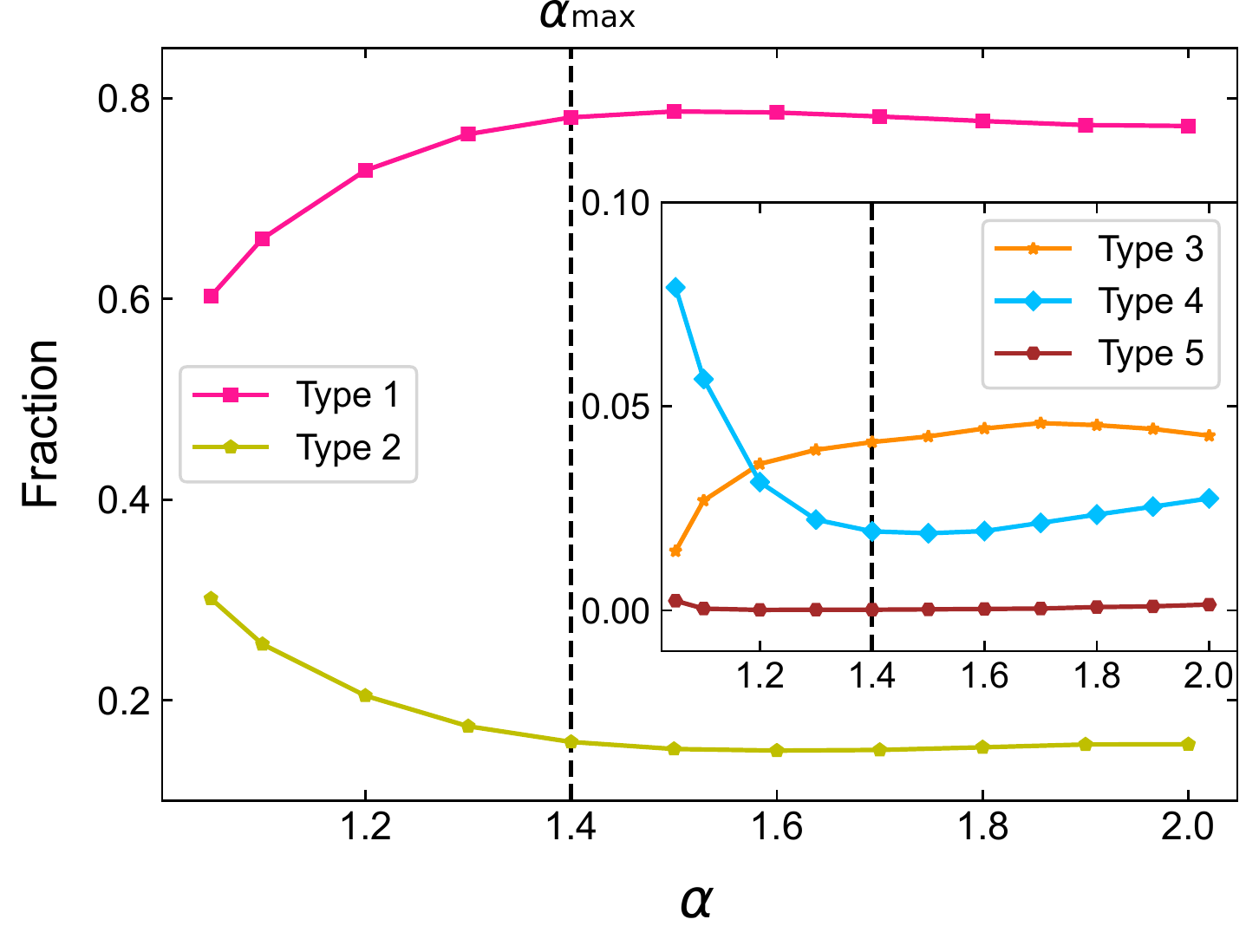}
	\caption{\label{Fig:Contacttypefraction}The fractions of the five contact configuration types of Table~\ref{table2} for packings of dimers with different $\alpha$.}
\end{figure}

In order to refine our analysis of the packing microstructure, we define five distinct contact configurations according to the number of contact points that are shared by two neighbouring dimers, see Table~\ref{table2}. Excluding the regime $\alpha\in [1,1.05)$, we determine how the fraction of each configuration type changes as a function of $\alpha$, see Fig.~\ref{Fig:Contacttypefraction}. We see that even though the average number of contacts $z$ is approximately constant over this range of $\alpha$, the underlying contact configurations change significantly with $\alpha$. Most notably, the two most common contact configurations, Type 1 and Type 2, increase and decrease, respectively, as $\alpha$ increases up to around $\alpha_{\rm max}$ and remain approximately unchanged for $\alpha>\alpha_{\rm max}$. The remaining contact configurations confirm this trend, showing the strongest variations in the regime $\alpha<\alpha_{\rm max}$. Overall, we see that contact configurations, in which spheres of neighbouring dimers only have one contact point (Type 1 and Type 3) increase, while those with multiple contact points (Types 2,4,5) decrease as the packing becomes denser up to the packing density peak at $\alpha_{\rm max}$. This trend is somewhat counter-intuitive, since the Type 2,4,5 configurations correspond to more optimal local arrangements between two dimers, which locally reduce the packing density. Similar results for the fractions of these five configuration types have been found for packings of shapes composed of four overlapping spheres \cite{Azema}.

Rather than excluding the aspect ratio regime where the problematic double and cusp contacts occur it might be tempting to re-assign such contacts and thus infer the properties of the small aspect ratio regime in an {\it ad-hoc} way. For example, a double contact as in Fig.~\ref{Fig:DCcontact}(a), which creates two overlaps of sphere pairs and is thus counted as two contact points, could be counted as only one, effectively ignoring the incorrect overlap with the interior sphere. This can be done likewise for other contact configurations, which require a careful consideration of the relative position and orientation of the overlapping dimer pair, see the full discussion in Appendix~\ref{Sec:map}. Re-assigning contacts in this way leads to a rapid but smooth decrease of $z$ to the corresponding value of spheres $z\approx 6$ as $\alpha\to 1$ (Fig.~\ref{Aspectratiocontactsmall}), but also exhibits seemingly unphysical behaviour, such as sharp peaks in the fractions of the Type 1--5 contact configurations around $\alpha\approx 1.05$, i.e., at the aspect ratio where double and cusp contacts start to occur (Fig.~\ref{Fig:Contacttypessmall}).

\subsection{Order metrics}

We employ several order metrics to measure global and local ordering in the dimer packings at various aspect ratios. The nematic orientational order parameter and the orientational pair correlation function are used to evaluate orientational ordering. Translational ordering is investigated with bond orientational order parameters, the radial distribution function and bond angle distributions. All calculations are made for the particles within the bulk volume so as to discard the crystallized region observed at the bottom of the container. 

\subsubsection{Metrics for orientational order}
\label{Sec:S}

The nematic orientational order parameter $S$ has traditionally been applied to identify different ordered phases of liquid crystals by characterising the average molecular orientation \cite{DeGennes1993}. $S$ is defined as:
\begin{equation}
\label{Sdef}
S=\langle P_2 (\cos\beta) \rangle\approx\frac{1}{N_{\rm b}} \sum_{i}^{N_{\rm b}}P_2 (\cos\beta_i)
\end{equation}
where $P_2(x)=\frac{1}{2}\left(3x^2-1\right)$ is the second Legendre polynomial and $\beta_i$ the angle between the orientation of dimer $i$ and the so-called director, which specifies the average orientation of the particles. The dimer orientation is described by the unit vector $\mathbf{u}^{(i)}$ measured along the dimer's long axis.

We apply this parameter to the dimer packings to quantify the global orientational order. When all $\mathbf{u}^{(i)}$ are randomly oriented, $S=0$, while if all $\mathbf{u}^{(i)}$ are oriented in a plane normal to the director, $S=-0.5$, which corresponds to a perfect oblate phase. When all $\mathbf{u}^{(i)}$ are aligned with the director, we have perfect nematic order with $S=1$.

In order to determine the director and $S$, we first evaluate the tensor $\Omega$ defined as:
\begin{equation}
	\Omega_{kl}=\frac{1}{N_{\rm b}} \sum_{i}^{N_{\rm b}}\left(\frac{3}{2}u^{(i)}_{k} u^{(i)}_{l}-\frac{1}{2}\delta_{kl}\right)
\end{equation}
Denoting by $\lambda_{\rm max}$ the eigenvalue of $\Omega$ with the largest absolute value, we identify the director as the eigenvector corresponding to $\lambda_{\rm max}$. For all aspect ratios, we find that the director is aligned with the $\hat{\mathbf{z}}$-axis (gravity direction). We then obtain $S$ directly as:
\begin{equation}
	S=\lambda_{\rm max}.
\end{equation}

We also determine the orientational pair correlation function $S_2$ in order to quantify local ordered structures at a radial distance $r$ from a reference particle. $S_2$ is calculated as:

\begin{equation}
\label{orients}
S_2(r)=\langle P_2 (\cos\beta_{ij})\delta(r-|\mathbf{r}_i-\mathbf{r}_j|) \rangle \approx \frac{ \sum\limits_{i=1}^{N_b}\sum\limits_{j \in n_i(r)}P_2 (\cos\beta_{ij}(r))}{\sum\limits_{i=1}^{N_b}|n_i(r)|}
\end{equation}
where $\cos\beta_{ij}=\mathbf{u}^{(i)}\cdot\mathbf{u}^{(j)}$ and $n_i(r)$ denotes the set of particles in a spherical shell of width $\Delta(r)=0.025d$ at a distance $r$ from the centre of dimer $i$ in the bulk. The expression $|n_i(r)|$ refers to the size (cardinality) of the set $n_i(r)$. We note that the spherical shell considered in $S_2$ can extend into the boundary region beyond the bulk and thus include particles in partially crystallized regions, although the effect on the average should be small. In general, due to the non-periodic boundary conditions in the $\hat{\mathbf{z}}$-direction our packings are not rotationally invariant and thus the restriction to a radial coordinate is only an approximation.

\begin{figure}
	\centering
	\includegraphics[width=0.45\textwidth]{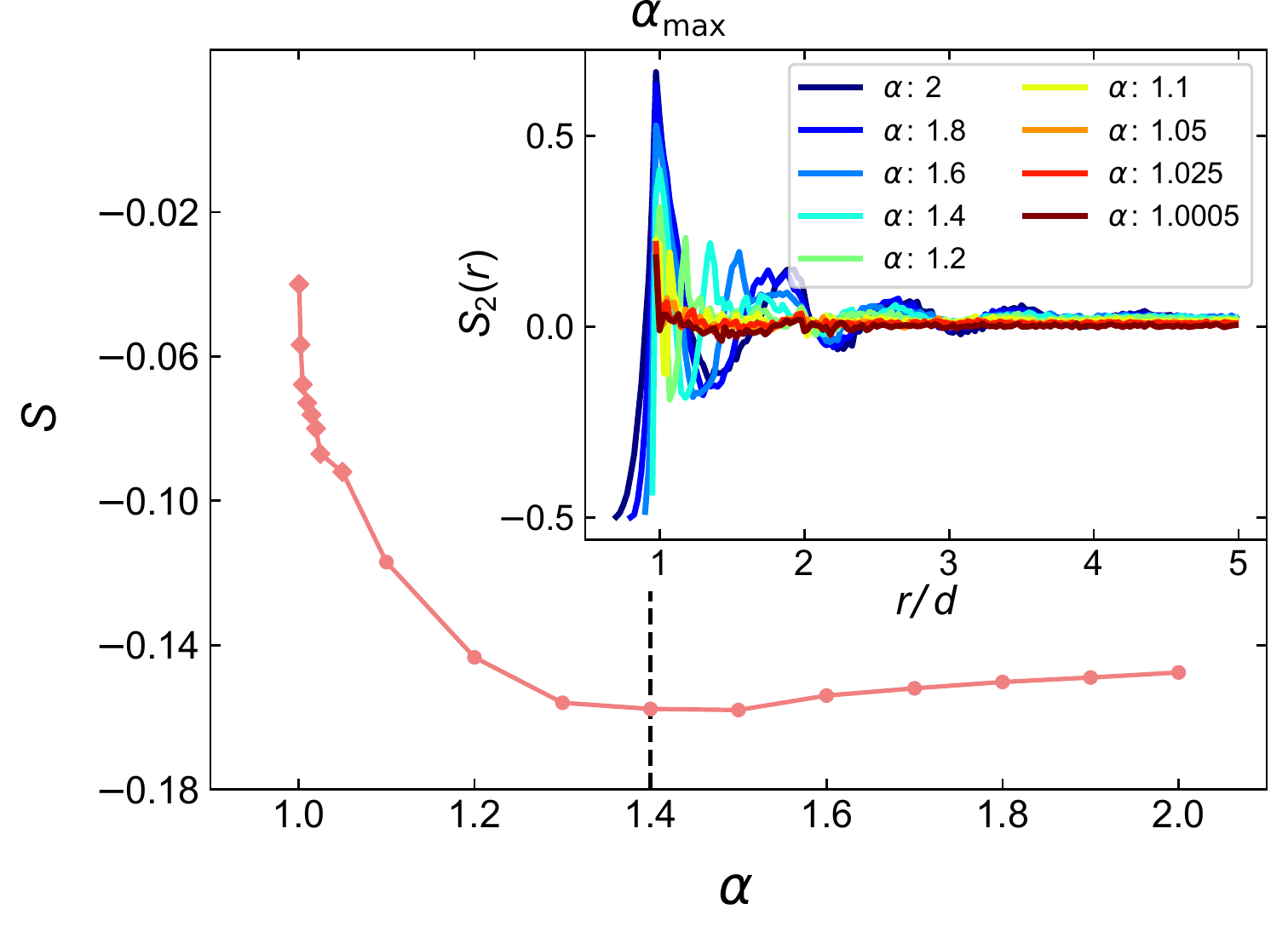}
	\caption{\label{Fig:Svsalpha}The nematic orientational order parameter $S$ vs the aspect ratio $\alpha$. Values of $S$ are shown averaged over 10 independent simulation runs for $\alpha\ge 1.1$ (dots), and for a single run for $\alpha<1.1$ (diamonds). Inset: the orientational pair correlation function $S_2$ vs $r/d$ for various aspect ratios.}
\end{figure}

We present the dependence of $S$ and $S_2(r)$ on the aspect ratio $\alpha$ in Fig.~\ref{Fig:Svsalpha}. We see that $S$ changes rapidly as $\alpha$ increases from the sphere value, reaching its minimum at around $\alpha_{\rm max}$ and remaining approximately constant for $\alpha>\alpha_{\rm max}$, in line with the behaviour of $z_{\rm c}$ and the different contact types. Interestingly, the behaviour of $S(\alpha)$ as $\alpha\to 1$ appears almost singular, but the range of values is not sufficient to identify a clear power-law. The minimum of $S$ at $\approx -0.16$ indicates slight oblate ordering, where the dimers' long axes are oriented close to the horizontal plane normal to the direction of gravity. This ordering is thus in agreement with that observed in simulation studies of prolate ellipsoids using also pouring under gravity \cite{Buchalter1994,Delaney2011,Gan2020}. In order to compare the magnitude of the orientational ordering with these studies, we also calculated the order parameter $\chi$ used in \cite{Buchalter1994,Delaney2011,Gan2020}, which is defined in Eq.~\eqref{chi}. We find a maximum of $\chi\approx 0.32$ for $\alpha=1.4$. By comparison, in \cite{Buchalter1994} the maximum is $\chi\approx 0.4$ for $\alpha\approx1.5$, while \cite{Delaney2011} and \cite{Gan2020} find $\chi\approx 0.25$ and $\chi\approx 0.5$, respectively, for $\alpha\approx1.5$. Note that in \cite{Delaney2011,Gan2020}, $\chi$ monotonically increases upon further elongation over the observed range of aspect ratios.

The plot of $S_2$ in Fig.~\ref{Fig:Svsalpha}(inset) demonstrates how orientational correlations become more long-range for larger aspect ratios. For small $\alpha$, correlations decay rapidly within the first coordination shell, while for large $\alpha$ oscillations in $S_2$ are visible over the whole range of $r/d$, which is here limited by $r/d=5$, i.e., the width of the boundary region on top of the bulk region that restricts the maximum radius of the spherical shell used in Eq.~\eqref{orients}.

\subsubsection{Bond orientational order parameters}

The bond-orientational order metrics $q_l$ and $Q_l$ introduced by Steinhardt \textit{et al}. \cite{Steinhardt1983} have most commonly been used to quantify translational order in disordered packings of spherical particles \cite{Kansal2002,Aste2005,Lochmann2006,Wouterse2006,Jin2010,Xu2010}. While $Q_l$ is widely accepted as a well-defined parameter to measure global ordering in a packing, it has been suggested that the local order parameter $q_l$ needs more caution to reliably identify local crystalline structures in these systems \cite{Kapfer2012,Mickel2013}. It was assumed that higher values of $q_6$ are associated with higher degrees of order \cite{Kansal2002} and averages $\langle q_6 \rangle$ have been used to quantify the overall degree of order for disordered sphere packings \cite{Lochmann2006}. However, it has been found that some local configurations of disordered sphere packings that are clearly non-crystalline have exhibited the same values of $q_6$ as hcp or fcc crystals \cite{Kapfer2012}. Therefore, in this study, we use recently introduced local order parameters defined by Eslami \textit{et al}. \cite{Eslami2018} to improve the accuracy of determining local translational order in the dimer packings.

\begin{figure}
\centering
\includegraphics[width=0.25\textwidth]{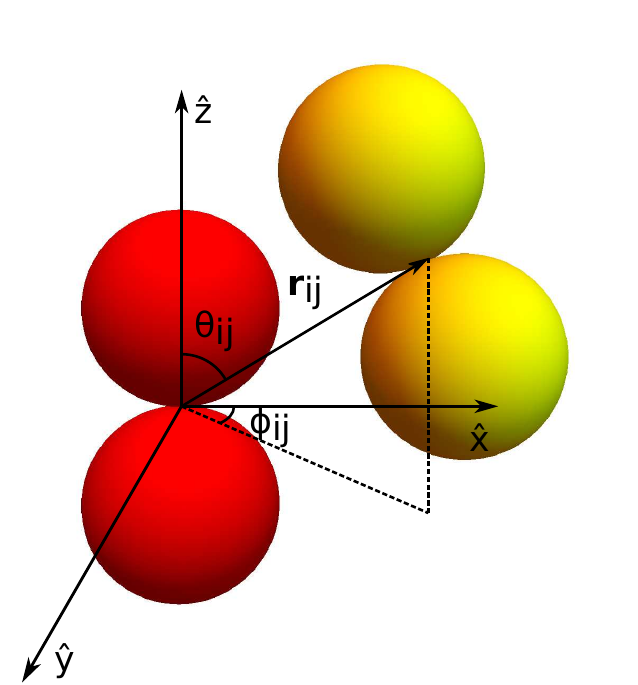}
\caption{\label{Fig:Bondillustration}Parametrization of the separation vector (bond vector) $\mathbf{r}_{ij}=\mathbf{r}_j-\mathbf{r}_i$ connecting the reference particle $i$ (red) with $j$ (yellow). The definitions of the polar and azimuthal angles, $\theta_{ij}$ and $\phi_{ij}$, respectively, are indicated.}
\end{figure}

Steinhardt \textit{et al}. \cite{Steinhardt1983} associated with every bond joining a particle and its neighbours a set of spherical harmonics:
\begin{equation}
	\label{qlm}
	q_{lm}(i)=\frac{1}{|NN(i)|} \sum_{j \in NN(i)} Y_{lm}(\theta_{ij}, \phi_{ij})
\end{equation}
where the $Y_{lm}$ are spherical harmonics and $\theta_{ij}$, $\phi_{ij}$ denote the polar and azimuthal angles which define the orientation of the vector (bond) pointing from the reference particle $i$ to another particle $j$, see Fig.~\ref{Fig:Bondillustration}. $NN(i)$ contains the set of neighbour indices for particle $i$, which are defined as those particles $j$ that have at least one contact with $i$.

The local orientational order parameter $q_l(i)$ of particle $i$ is then defined as the following rotational invariant combination of $q_{lm}$:
\begin{equation}
q_l(i)=\sqrt{\frac{4\pi}{2l+1} \sum_{m=-l}^{l} {|q_{lm}(i)|}^2}.
\end{equation}
Moreover, the global orientational order parameter $\mathit{Q_l}$ is defined as
\begin{equation}
Q_l={\left(\frac{4\pi}{2l+1}\sum_{m=-l}^{l} |Q_{lm}|^2\right)}^{1/2},
\end{equation}
where 
\begin{equation}
Q_{lm}=\frac{1}{N_{\rm b}} \sum_{i=1}^{N_{\rm b}} {q_{lm}(i)}
\end{equation}

Recently, Eslami \textit{et al.} introduced the local order parameters $\bar{\tilde{q}}_{l}(i)$ to improve the determination of liquid and different crystallized phases \cite{Eslami2018}. Starting from the $q_{lm}$ of Eq.~\eqref{qlm}, we first determine
\begin{equation}
	\tilde{q}_{l}(i)= \frac{1}{|NN(i)|} \sum_{j \in NN(i)} \sum_{m=-l}^{l} {\hat{q}_{lm}(i) {\hat{q}_{lm}}^\ast(j)}
\end{equation}
where $\hat{q}_{lm}^\ast(j)$ is the complex conjugate of $\hat{q}_{lm}(j)$ and $\hat{q}_{lm}(i)$ is defined as follows:
\begin{equation}
	\hat{q}_{lm}(i)=\frac{q_{lm}(i)}{{\left(\sum\limits_{m=-l}^{l} {|q_{lm}(i)|^2} \right)}^{1/2}}
\end{equation}
Then the order parameters $\bar{\tilde{q}}_{l}(i)$ are obtained by averaging over the first coordination shell of particle $i$:
\begin{equation}
\label{eslami1}
	\bar{\tilde{q}}_{l}(i)=\frac{1}{1+|NN(i)|}  \left[\tilde{q}_{l}(i)+ \sum_{j \in NN(i)} {\tilde{q}_{l}(j)}  \right]
\end{equation}

\begin{figure}
	\includegraphics[width=0.22\textwidth]{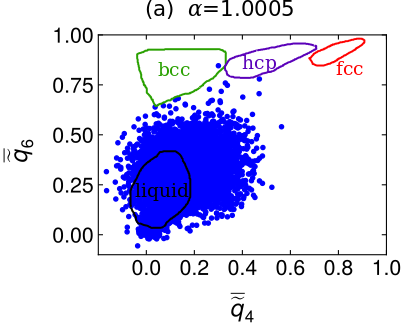}
	\includegraphics[width=0.22\textwidth]{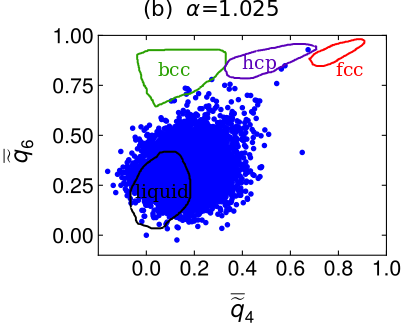}
	\includegraphics[width=0.22\textwidth]{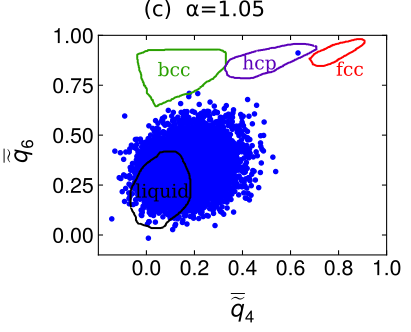}
	\includegraphics[width=0.22\textwidth]{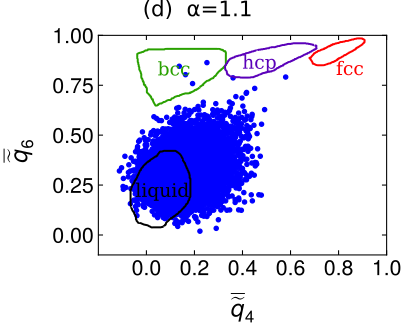}
	\includegraphics[width=0.22\textwidth]{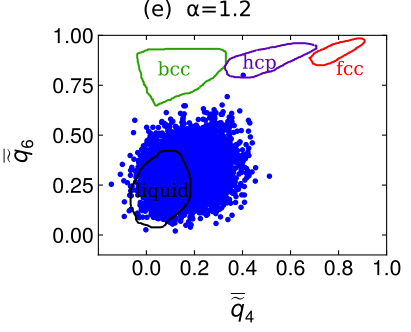}
	\includegraphics[width=0.22\textwidth]{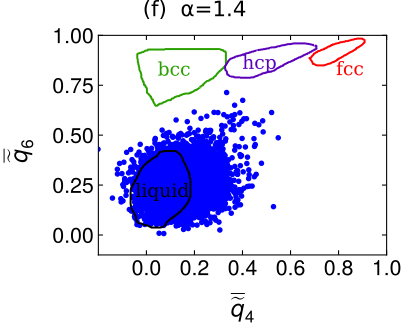}
	\includegraphics[width=0.22\textwidth]{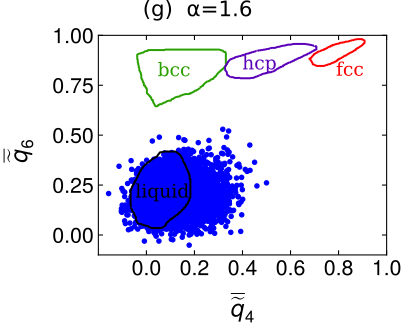}
	\includegraphics[width=0.22\textwidth]{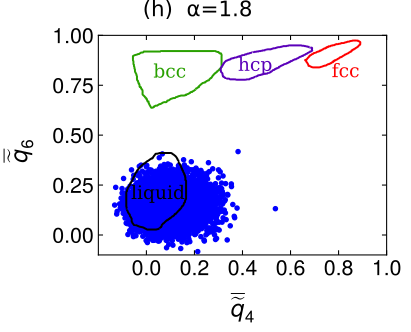}
	\includegraphics[width=0.22\textwidth]{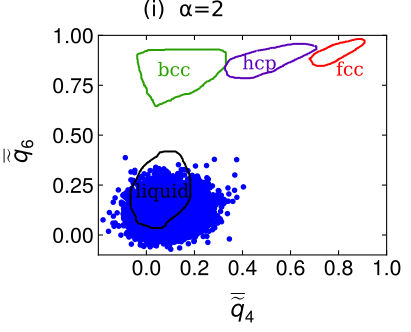}
	\caption{\label{Fig:q4andq6}The local order parameters $\overline{\widetilde{q}}_{4}$ and $\overline{\widetilde{q}}_{6}$ defined in Eq.~\eqref{eslami1}. Every data point corresponds to a dimer in the bulk region of the packing. The sketched regions for bcc, hcp, fcc, and liquid phases of Lennard-Jones particles are taken from \cite{Eslami2018}.}
\end{figure} 

The advantage of $\bar{\tilde{q}}_{l}(i)$ over $q_l$ is that they can distinguish the liquid phase and different crystalline phases in a more accurate way \cite{Eslami2018}. They indicate in fact the correlation between the order in the first and the second coordination shell of a reference particle \cite{Eslami2018}. It has been observed that $\bar{\tilde{q}}_{6}(i)$ is large $\approx1$ for crystalline phases, while $\bar{\tilde{q}}_{6}(i)$ assumes values close to zero for disordered (liquid) phases, which thus allows to easily discriminate between such phases. On the other hand, the values of $\bar{\tilde{q}}_{4}(i)$ are sensitive to the crystal type, so $\bar{\tilde{q}}_{4}(i)$ is able to distinguish bcc, fcc, and hcp crystals.

We display the pairs ($\overline{\widetilde{q}}_{4}$,$\overline{\widetilde{q}}_{6}$) for each dimer in the bulk region of the packing in Fig.~\ref{Fig:q4andq6} for various aspect ratios. By comparing these results to empirical data for liquid, bcc, hcp, and fcc phases of Lennard-Jones particles from \cite{Eslami2018}, we observe that the distributions at large aspect ratios ($\alpha>1.4$) are quite clearly in a liquid phase where $-0.05<\overline{\widetilde{q}}_{4}<0.3$ and $0<\overline{\widetilde{q}}_{6}<0.4$. As the aspect ratio decreases, the region occupied by $\overline{\widetilde{q}}_{4}$ and $\overline{\widetilde{q}}_{6}$ expands and approaches the region occupied by the bcc/hcp crystal phases indicating the presence of a large proportion of dimers exhibiting some local translational order intermediate between a liquid and bcc/hcp crystalline order.

\begin{figure}
	\centering	
	\includegraphics[width=0.45\textwidth]{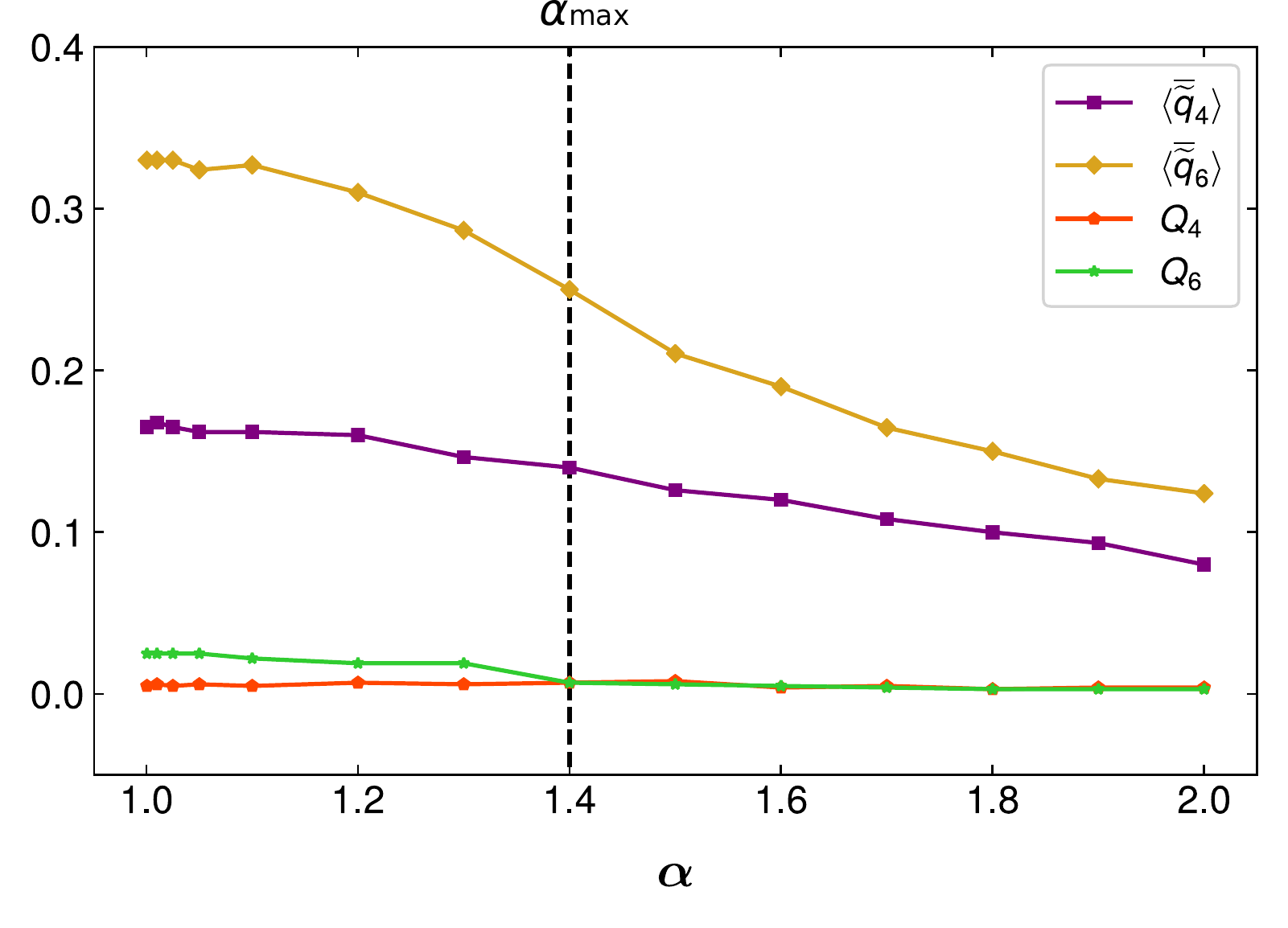}
	\caption{\label{Fig:BOmean}The global bond orientational order parameters $Q_4$, $Q_6$ and the averages $\langle\overline{\widetilde{q}}_{4}\rangle$ and $\langle\overline{\widetilde{q}}_{6}\rangle$ vs $\alpha$. By comparison, $\langle\overline{\widetilde{q}}_{4}\rangle\approx0.06$ and $\langle\overline{\widetilde{q}}_{6}\rangle\approx0.2$ for the liquid phase of Lennard-Jones particles \cite{Eslami2018}.}
\end{figure}

We also calculate the averages $\langle\overline{\widetilde{q}}_{4}\rangle$, $\langle\overline{\widetilde{q}}_{6}\rangle$ and compare their values with the global order parameters $Q_4$, $Q_6$ for different aspect ratios, see Fig.~\ref{Fig:BOmean}. While $Q_4$ is close to zero for all aspect ratios, there is a slight increase in $Q_6$ for $\alpha<1.4$ implying some global ordering at small aspect ratios. In line with the observations in Fig.~\ref{Fig:q4andq6}, we see that both $\langle\overline{\widetilde{q}}_{4}\rangle$ and $\langle\overline{\widetilde{q}}_{6}\rangle$ are non-zero and monotonically decreasing as $\alpha$ increases, whereby $\langle\overline{\widetilde{q}}_{6}\rangle$ varies over a larger range than $\langle\overline{\widetilde{q}}_{4}\rangle$. For small aspect ratios, both averages are considerably larger than the corresponding averages of a fluid phase, which were determined as $\langle\overline{\widetilde{q}}_{4}\rangle\approx0.06$ and $\langle\overline{\widetilde{q}}_{6}\rangle\approx0.2$. Overall, we observe that at large aspect ratios the packing is more translationally disordered than at small aspect ratios.

\subsubsection{Radial distribution function}
\begin{figure}
	\centering	
	\includegraphics[width=0.45\textwidth]{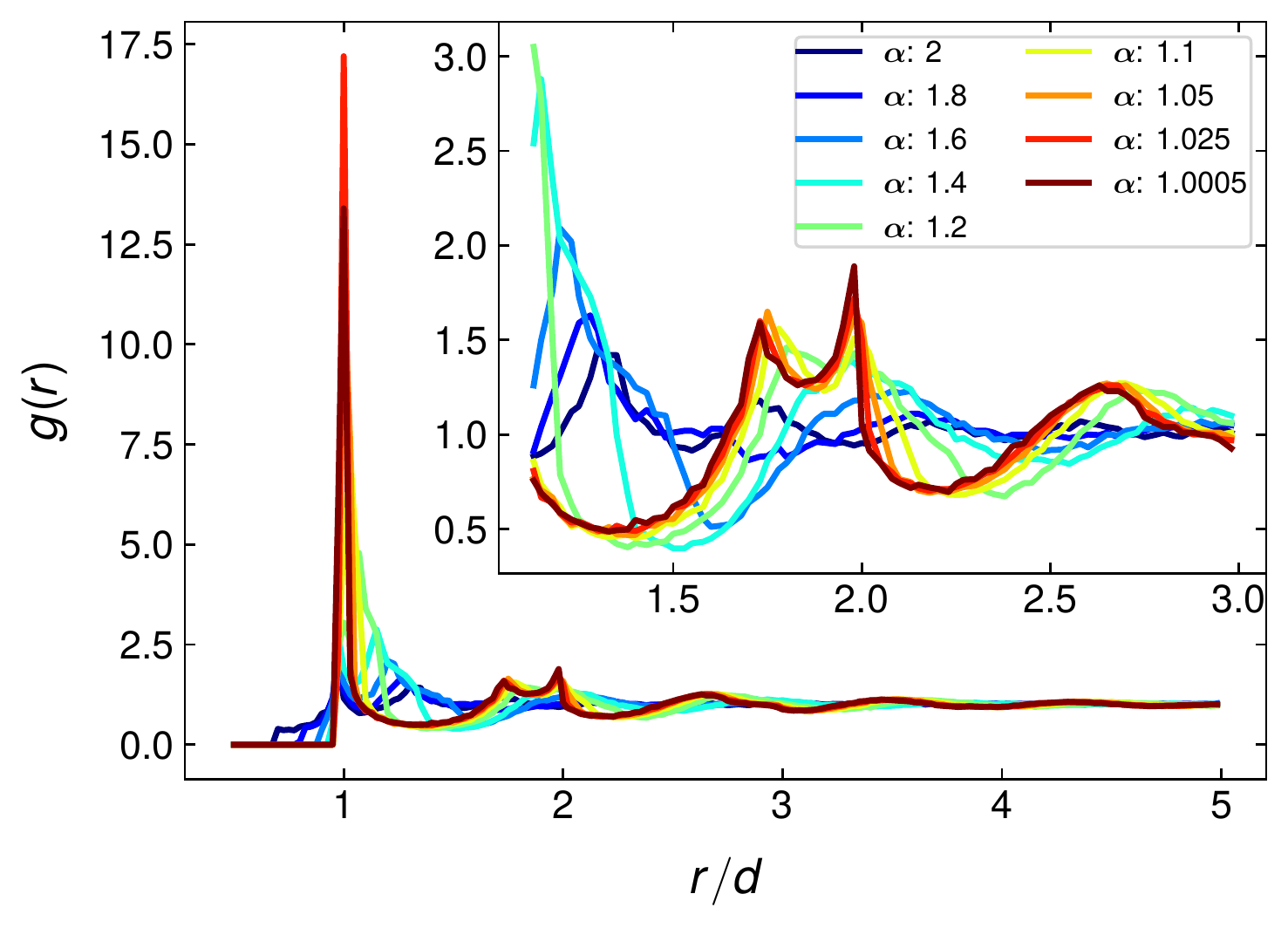}	
	\caption{\label{Fig:Rdf} The radial distribution function $g(r)$ of the dimer packings, Eq.~\eqref{gr}, for different $\alpha$. Inset: enlargement of the regime $r/d\in[1.125,3]$.}
\end{figure}

We calculate the radial distribution function $g(r)$ to further examine the translational correlations between the dimers. The radial distribution function of the bulk dimers is determined as 

\begin{equation}
\label{gr}
g(r)=\frac{\sum\limits_{i=1}^{N_{\rm b}}|n_i(r)|}{N_{\rm b} \rho V_{\rm shell}(r)}
\end{equation}
where $n_i(r)$ denotes the set of particles in a spherical shell of width $\Delta(r)=0.025d$ at a distance $r$ from the centre of dimer $i$ in the bulk, $\rho$ is the particle number density, and $V_{\rm shell}(r)$ is the volume of the shell. As discussed for the orientational correlation function $S_2(r)$, Eq.~\eqref{orients}, the restriction to a radial coordinate is only an approximation due to the fact the our packings are not rotationally invariant. As before the spherical shell can extend into the boundary region beyond the bulk. We plot $g(r)$ as a function of $r/d$ for various aspect ratios in Fig.~\ref{Fig:Rdf}. We see that for small aspect ratios $g(r)$ exhibits the characteristic shape of sphere packings with a main peak at $r/d=1$ and a split second-peak at $r/d\approx1.7$ and $r/d\approx2$ \cite{Silbert2002,Williams2003,Donev2005,Wouterse2007,Zhao2012}. For larger aspect ratios, these sharp peaks broaden and reduce in height. These results are consistent with the variation of bond orientational correlations with the aspect ratio discussed above, where elongation in the dimers results in a reduction of translational correlations.

\subsubsection{Bond angle distribution}

\begin{figure*}
	\centering 	
	\includegraphics[width=0.2\textwidth]{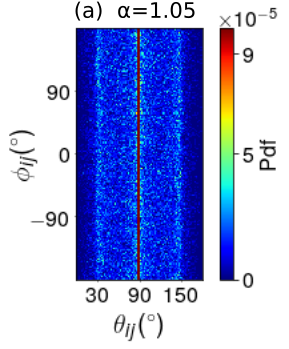}\hspace{0.5cm}
	\includegraphics[width=0.2\textwidth]{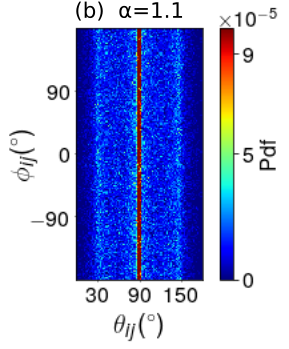}\hspace{0.5cm}
	\includegraphics[width=0.2\textwidth]{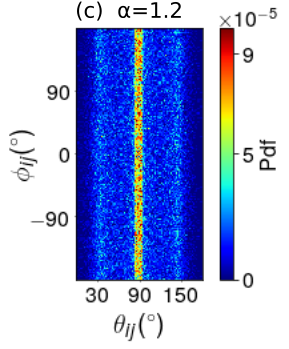}\hspace{0.5cm}
	\includegraphics[width=0.2\textwidth]{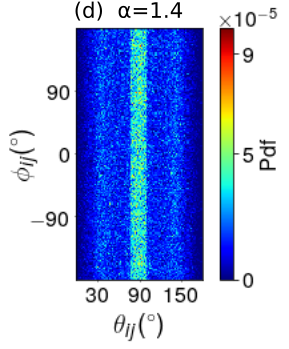}\vspace{0.5cm}
	\includegraphics[width=0.2\textwidth]{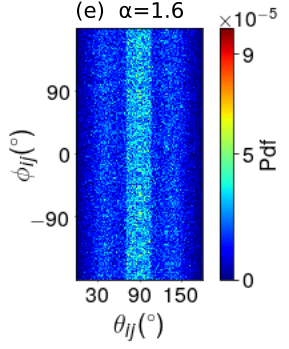}\hspace{0.5cm}
	\includegraphics[width=0.2\textwidth]{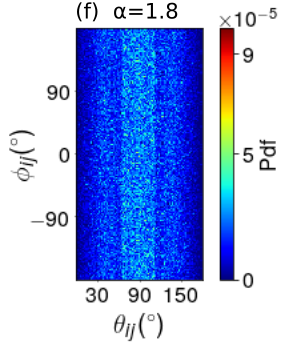}\hspace{0.5cm}
	\includegraphics[width=0.2\textwidth]{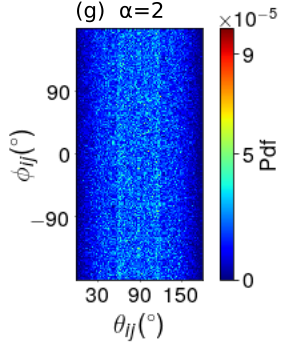}
	\caption{\label{Fig:Bondangledist}PDFs of the polar and azimuthal angles $\theta_{ij}, \phi_{ij}$ of the bond vectors $\mathbf{r}_{ij}$ for all neighbour pairs $i,j$ and different aspect ratios.}
\end{figure*}

\begin{figure*}
	\centering 	
	\includegraphics[width=0.18\textwidth]{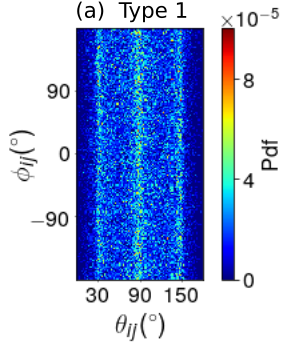}\hspace{0.25cm}
	\includegraphics[width=0.18\textwidth]{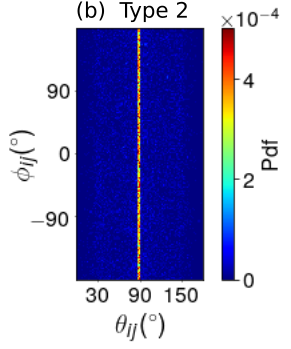}\hspace{0.25cm}
	\includegraphics[width=0.18\textwidth]{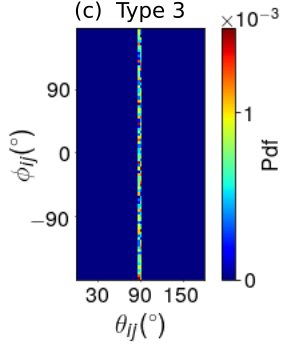}\hspace{0.25cm}
	\includegraphics[width=0.18\textwidth]{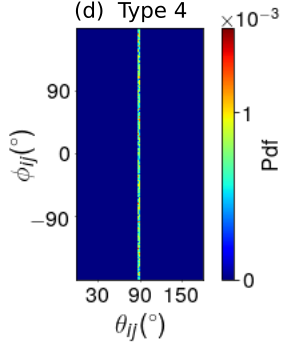}\hspace{0.25cm}
	\includegraphics[width=0.18\textwidth]{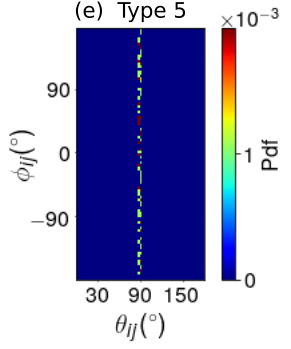}
	\caption{\label{Fig:Bondangledist1}PDFs of the polar and azimuthal angles $\theta_{ij}, \phi_{ij}$ of the bond vectors $\mathbf{r}_{ij}$ for all neighbour pairs $i,j$ with a specific contact type. Aspect ratio: $\alpha=1.05$.}
\end{figure*}
\begin{figure*}
	\centering 	
	\includegraphics[width=0.18\textwidth]{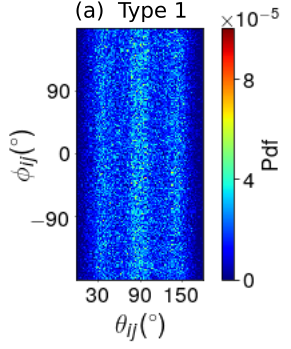}\hspace{0.25cm}
	\includegraphics[width=0.18\textwidth]{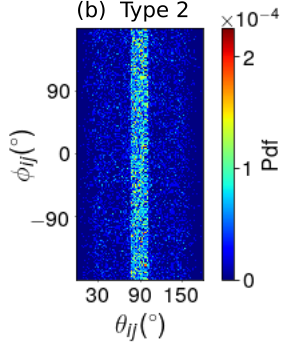}\hspace{0.25cm}
	\includegraphics[width=0.18\textwidth]{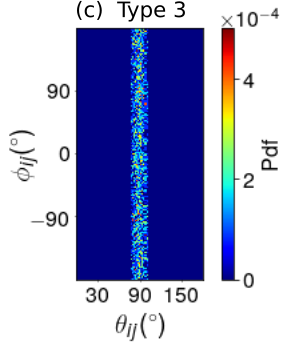}\hspace{0.25cm}
	\includegraphics[width=0.18\textwidth]{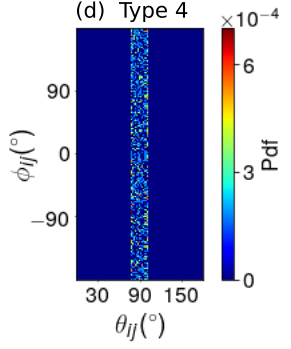}\hspace{0.25cm}
	\includegraphics[width=0.18\textwidth]{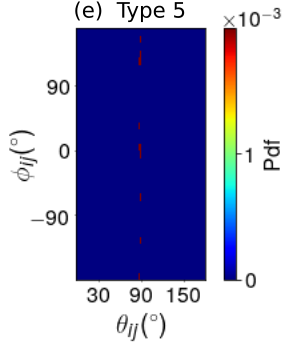}
	\caption{\label{Fig:Bondangledist2}PDFs of the polar and azimuthal angles $\theta_{ij}, \phi_{ij}$ of the bond vectors $\mathbf{r}_{ij}$ for all neighbour pairs $i,j$ with a specific contact type. Aspect ratio: $\alpha=\alpha_{\rm max}=1.4$.}
\end{figure*}
\begin{figure*}
	\centering 	
	\includegraphics[width=0.18\textwidth]{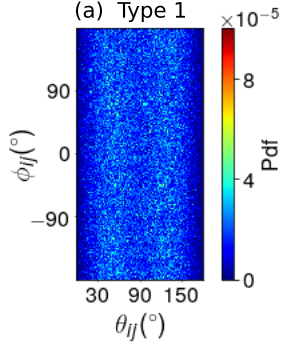}\hspace{0.25cm}
	\includegraphics[width=0.18\textwidth]{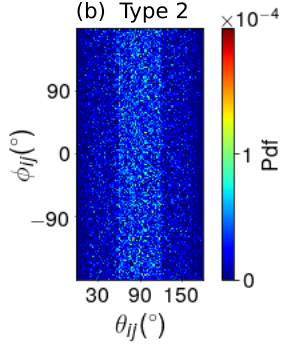}\hspace{0.25cm}	
	\includegraphics[width=0.18\textwidth]{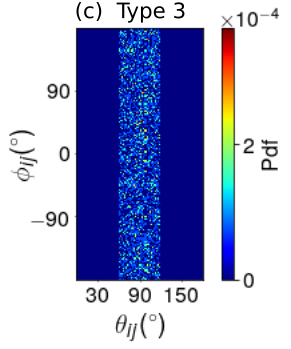}\hspace{0.25cm}
	\includegraphics[width=0.18\textwidth]{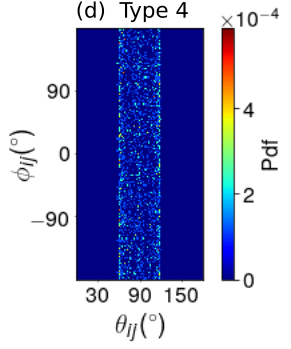}\hspace{0.25cm}
	\includegraphics[width=0.18\textwidth]{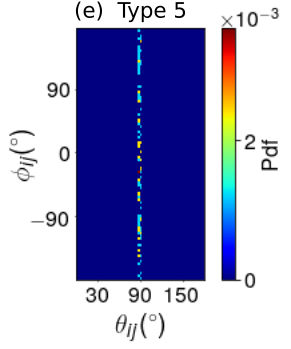}\hspace{0.25cm}
	\caption{\label{Fig:Bondangledist3}PDFs of the polar and azimuthal angles $\theta_{ij}, \phi_{ij}$ of the bond vectors $\mathbf{r}_{ij}$ for all neighbour pairs $i,j$ with a specific contact type. Aspect ratio: $\alpha=2$.}
\end{figure*}

We measure the probability for a dimer to have a contact at a particular direction relative to its long axis. For each dimer pair $i,j$, we determine the polar angle $\theta_{ij}$ and the azimuthal angle $\phi_{ij}$ of the bond vector $\mathbf{r}_{ij}=\mathbf{r}_j-\mathbf{r}_i$ in the reference frame of particle $i$, see Fig.~\ref{Fig:Bondillustration}. The probability density functions (PDFs) of $\theta_{ij}$ and $\phi_{ij}$ are shown for various aspect ratios in Fig.~\ref{Fig:Bondangledist}. It can be clearly seen from Fig.~\ref{Fig:Bondangledist} that at small aspect ratios dimers have primarily contacts at $\theta_{ij}=90^{\circ}$. As the aspect ratio increases, the band around $90^{\circ}$ widens and finally almost disappears at $\alpha=2$. For small aspect ratios, there are also symmetric secondary peaks visible at $\theta_{ij}=30^{\circ}$ and $\theta_{ij}=150^{\circ}$, with all contacts occurring within the range $\theta_{ij}\in[30^{\circ},150^{\circ}]$ up to $\alpha\approx 1.4$.

To get a better insight into the origin of these structures, the PDFs of $\theta_{ij},\phi_{ij}$ are further refined according to the contact configuration type between neighbouring dimers, see Figs.~\ref{Fig:Bondangledist1}--\ref{Fig:Bondangledist3}. For aspect ratio $\alpha=1.05$ (Fig.~\ref{Fig:Bondangledist1}), we see that for Type 2---5 only configurations with $\theta_{ij}\approx90^{\circ}$ are possible due to the geometric constraint of these configuration types. The structure observed in the overall bond diagram at very small aspect ratios (Fig.~\ref{Fig:Bondangledist}a and b) is thus primarily due to Type 1 configurations and the peak at $\theta_{ij}\approx90^{\circ}$. For larger aspect ratios $\alpha=\alpha_{\rm max}=1.4$ and $\alpha=2$, the bands for Type 2---4 widen due to the increase in possible relative orientations that still satisfy the contact constraint (see Figs.~\ref{Fig:Bondangledist2} and \ref{Fig:Bondangledist3}). This excludes Type 5 configurations which are available only in a narrow width of possible polar angles by definition. As expected, Type 1 configurations with only a single contact point between neighbours, which thus least constrains the relative orientations, exhibit a wide band of possible polar angles at all aspect ratios, see Figs.~\ref{Fig:Bondangledist1}(a),\ref{Fig:Bondangledist2}(a),\ref{Fig:Bondangledist3}(a). Interestingly, this band still exhibits some structure, with a main peak at $\theta_{ij}=90^{\circ}$ and symmetric secondary peaks at $\theta_{ij}=30^{\circ}$ and $\theta_{ij}=150^{\circ}$ for both $\alpha=1.05$ and $\alpha=1.4$, which disappear for $\alpha=2$.

\section{Conclusions}
\label{Sec:conclusion}

One of the main results of our study is the identification of structural features that accompany the formation of the peak in the packing density of elongated non-spherical particles. In particular, we find that (i) the coordination number $z_{\rm c}$; (ii) the fractions of Type 1--4 contact configurations; and (iii) the nematic order parameter $S$ undergo rapid changes upon deforming spheres into dimers with aspect ratios up to $\alpha\approx\alpha_{\rm max}$, while further elongation of the dimers leaves these metrics largely unchanged. This highlights that the peak in the packing density of Fig.~\ref{Fig:phi} arises due to microscopic re-arrangements up to $\alpha\approx\alpha_{\rm max}$ and subsequent excluded volume effects: the contact configurations remain statistically unchanged for $\alpha>\alpha_{\rm max}$, but since the particles are longer the packing can sustain more empty space while being mechanically stable, in line with the phenomenological description of spherocylinder packings using the random contact equation, which predicts a decay $\phi_{\rm j}\sim 1/\alpha$ \cite{Philipse1996}.

Dimers are a convenient shape model, because their contact interactions can be easily implemented by overlapping spheres. As such they represent one of the simplest non-spherical and concave shapes. However, our analysis shows that such a particle model does not allow to resolve the contact configurations at very small aspect ratios when interactions are not truly hard. As such we are not able to probe in our simulations, e.g., the analytical predictions from effective medium theory on the contact number scaling for very small shape deformations \cite{Brito2018}. The problematic double and cusp contacts should generally occur for shapes composed of overlapping (soft) spheres as used, e.g., in the optimization studies of \cite{Miskin2014,Roth2016}, which might prevent a detailed analysis of the contact properties of such simulated packings.

Our investigation highlights the competition between orientational and translational correlations between particles as a result of elongation. While the translational correlations are larger for small aspect ratios, the elongation induces the dimers to have both more orientationally ordered local structures (with slight global oblate ordering) and less translational order akin to those of a liquid. Dimers at large aspect ratios thus exhibit structures that resemble a liquid crystal in terms of these metrics. Importantly, the structural features identified here might be specific to the gravitational packing protocol used and might not occur in dimer packings obtained with other packing methods such as energy minimization from a random initial configuration \cite{Shiraishi2020}. Nevertheless, due to the simplicity of the protocol, which is also relevant in many real world scenarios, we expect our results to be significant to understand the packing density and structural properties of real granular matter composed of non-spherical particles.


\section*{Conflicts of interest}
There are no conflicts to declare.

\section*{Acknowledgements}
This research utilised Queen Mary's Apocrita HPC facility, supported by QMUL Research-IT http://doi.org/10.5281/zenodo.438045. We acknowledge the assistance of the ITS Research team at Queen Mary University of London.





\bibliographystyle{rsc} 

\begin{appendix}
	\section{Calculation of the dimer volume}
	\label{App:overlap}
\begin{figure}[h]
		\centering
		\includegraphics[width=0.225\textwidth]{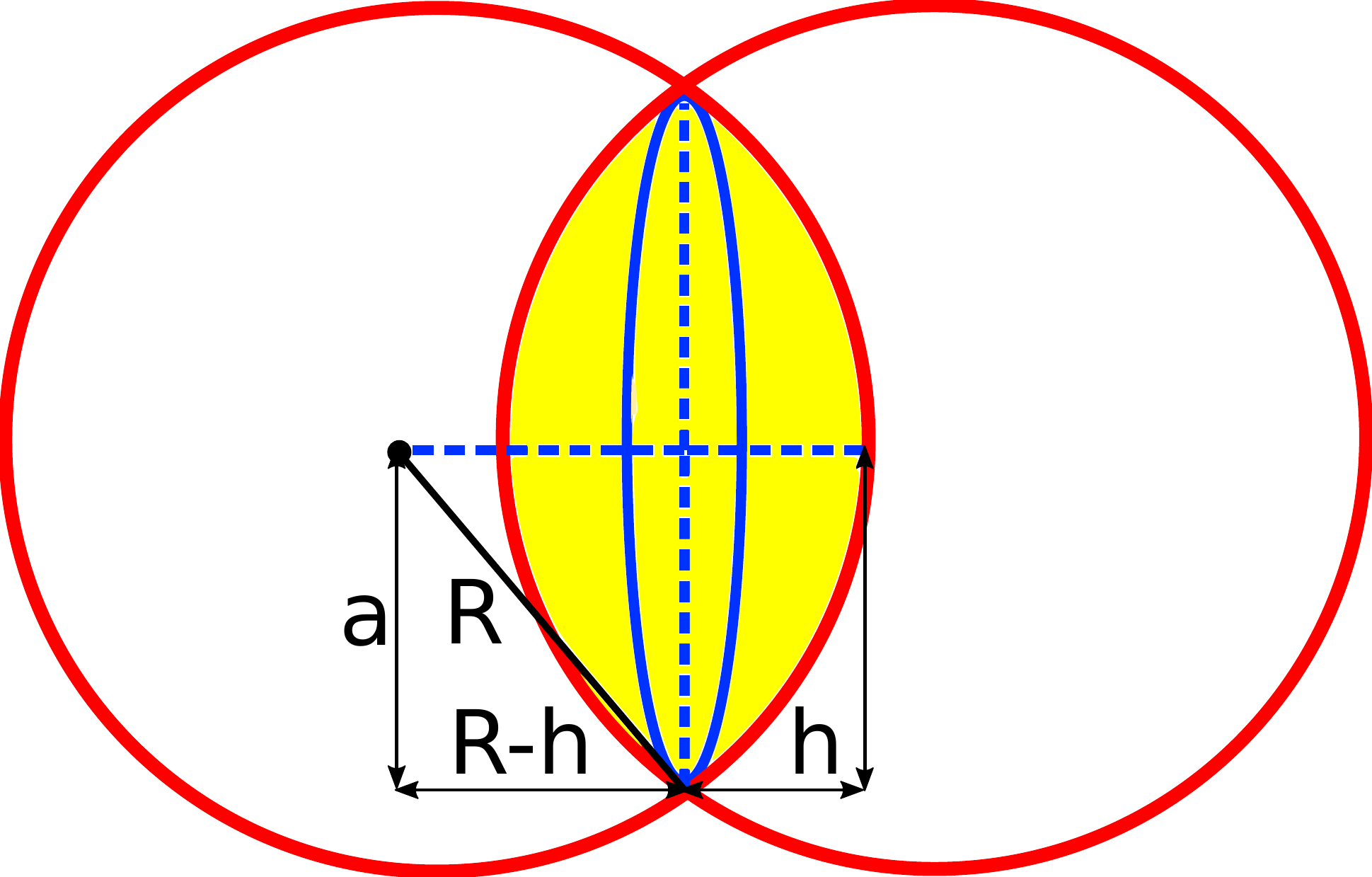}
		\caption{\label{Overlapvolume} The overlap volume of a dimer contains two equal spherical caps of height $h$ (coloured in yellow).} 
	\end{figure}

	The overlap volume of the two constituent spheres of a dimer contains two equal spherical caps which lie above/below the plane through the cusp points at the dimer's centre, see Fig.~\ref{Overlapvolume}. The volume of a spherical cap $V_{\rm cap}$ of height $h$ is found as:
	\begin{equation}
		V_{\rm cap}=\frac{1}{3}\pi h^2 (3R-h)
	\end{equation}
	where $R$ is the sphere radius.
	The dimer volume $V_{\alpha}$ is then calculated by subtracting the overlap volume from the sum of its constituent sphere volumes $V_{\rm sphere}=\frac{4}{3}\pi\,R^3$ as:
	\begin{equation}
		V_{\alpha}=2V_{\rm sphere}-2V_{\rm cap}
	\end{equation}
	
	\section{Algorithm for the identification of double and cusp contacts}	
	
\begin{figure}[h]
		\centering
		(a)	\includegraphics[width=0.125\textwidth]{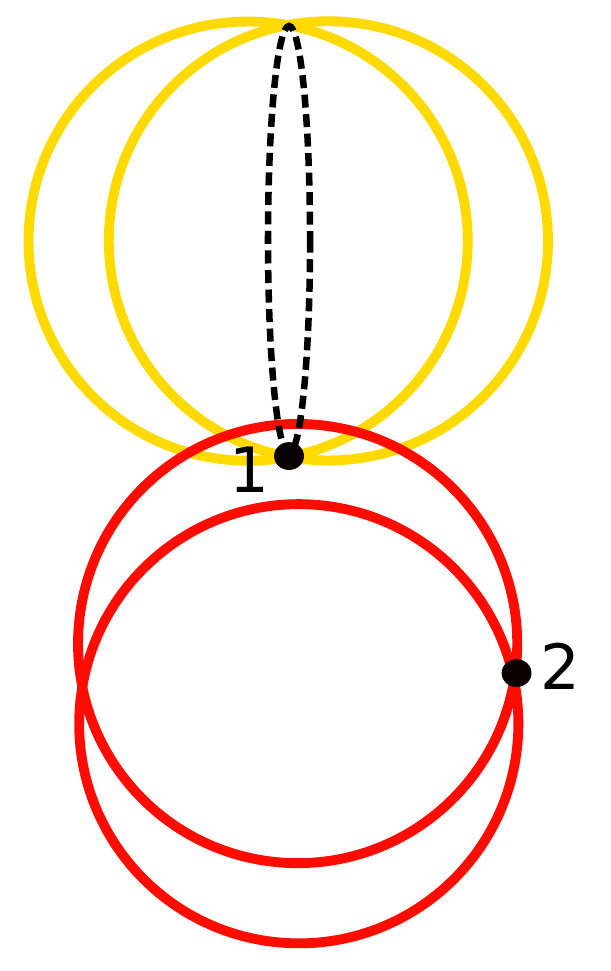} \hspace{0.5cm}
		(b)	\includegraphics[width=0.2\textwidth]{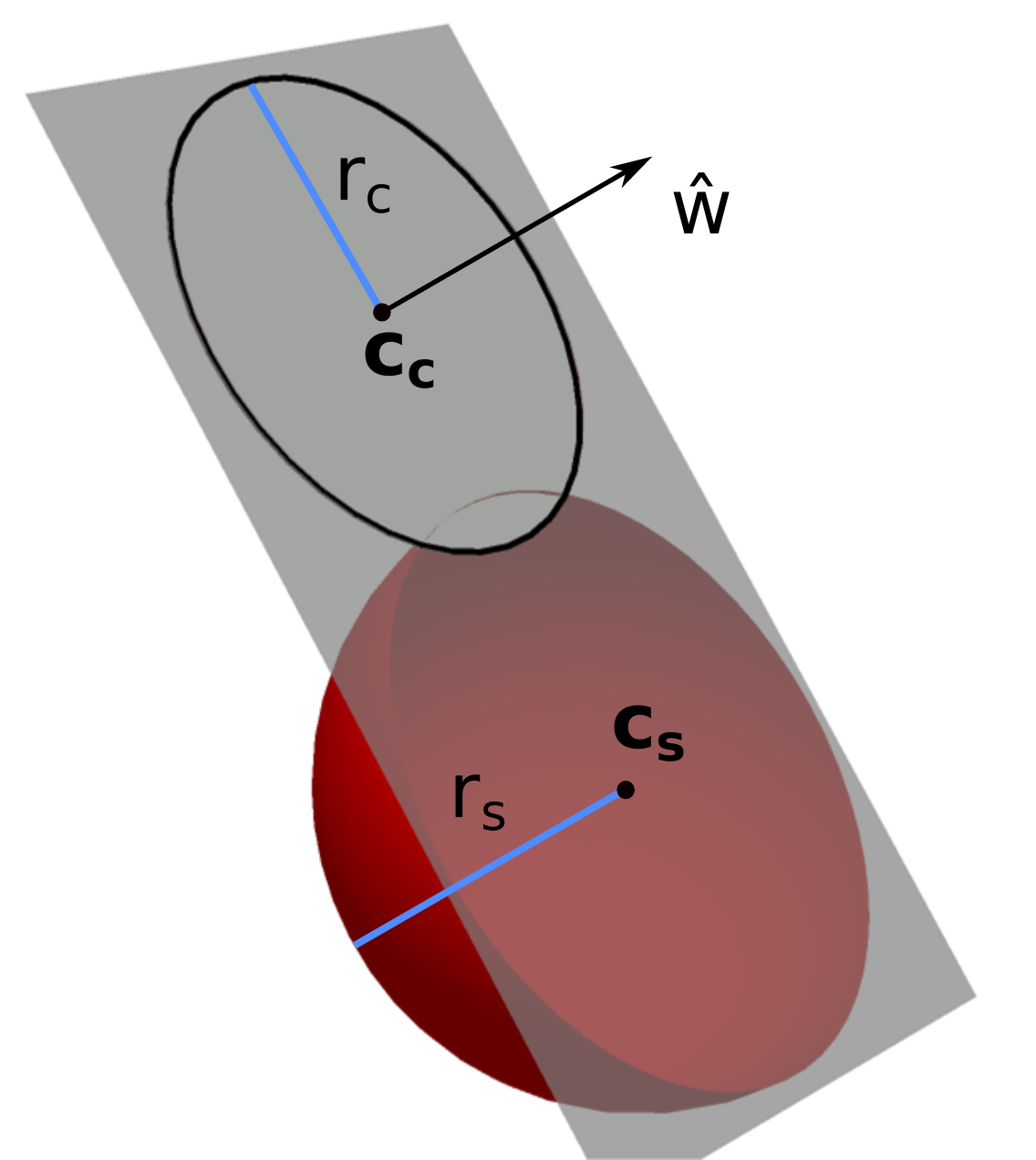} \hspace{0.5cm}\\
		\vspace{0.5cm}
		(c)	\includegraphics[width=0.2\textwidth]{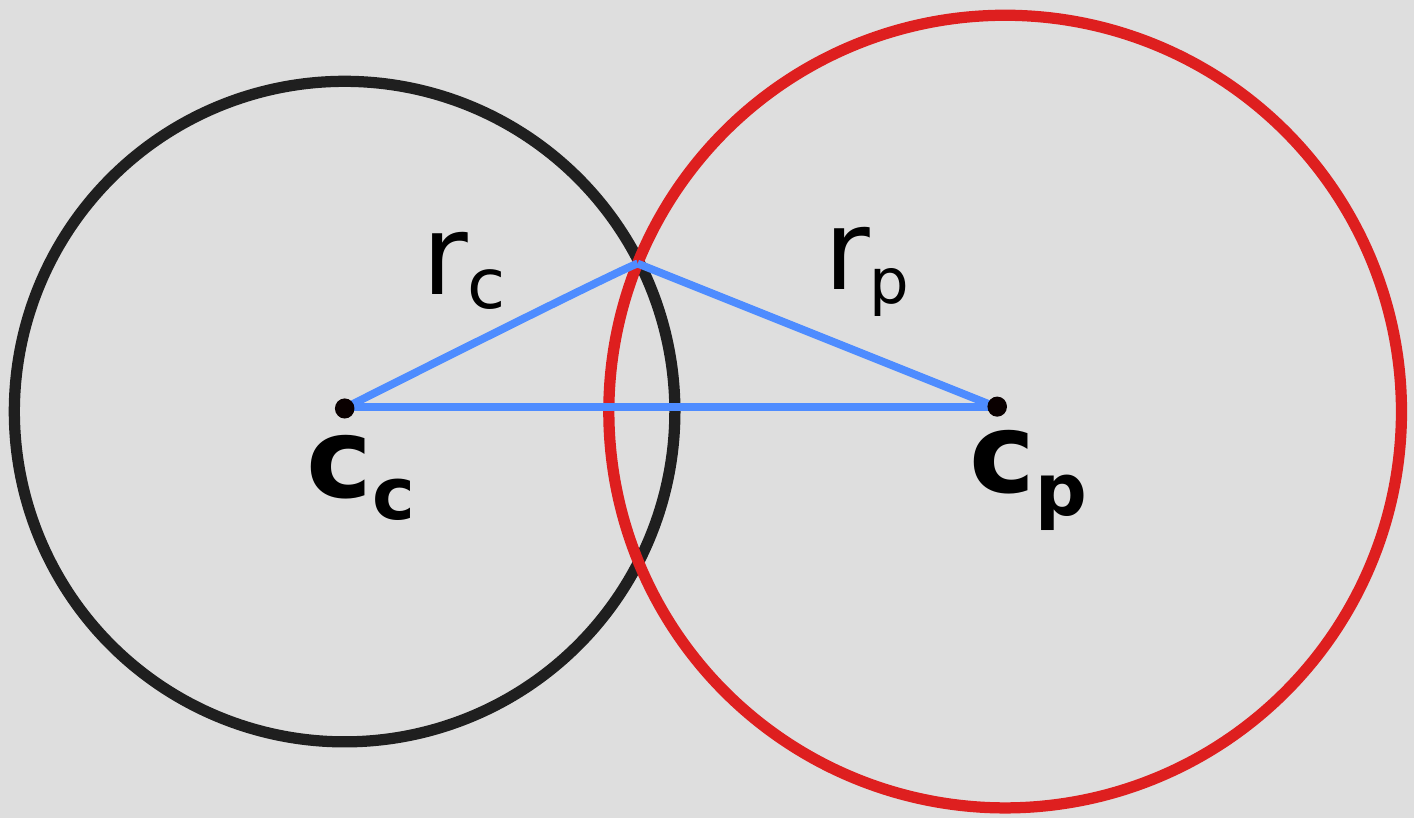}
		\caption{\label{Fig:DCcontactdetection1}Detecting double and cusp contacts. (a) First, it is checked if there is any overlap between the black circle (dashed) enclosing cusp located on the yellow dimer's surface and the contacting sphere of the red dimer. If there is an overlap between the circle and the sphere, it is identified as a cusp contact. (b) 3D Visualization of the circle and sphere interaction, it is determined if the plane of the circle cuts the sphere or not. (c) If the plane of the circle cuts the sphere, it forms a new circle (red) and then it is checked if there is overlap between the original circle and the new red circle.}
	\end{figure}

\begin{figure}[h]
		\centering
		(a)\includegraphics[width=0.2\textwidth]{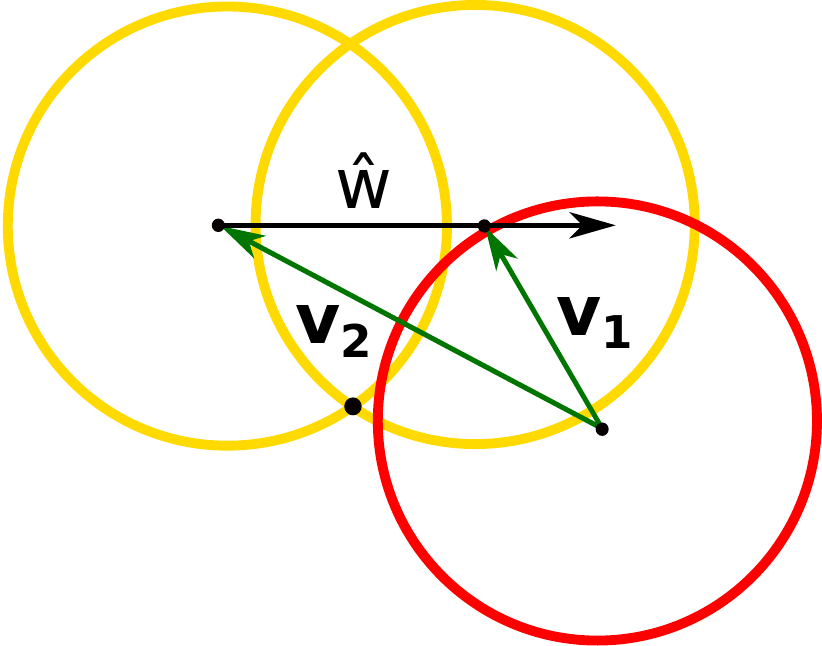}\hspace{0.25cm}
		(b)\includegraphics[width=0.2\textwidth]{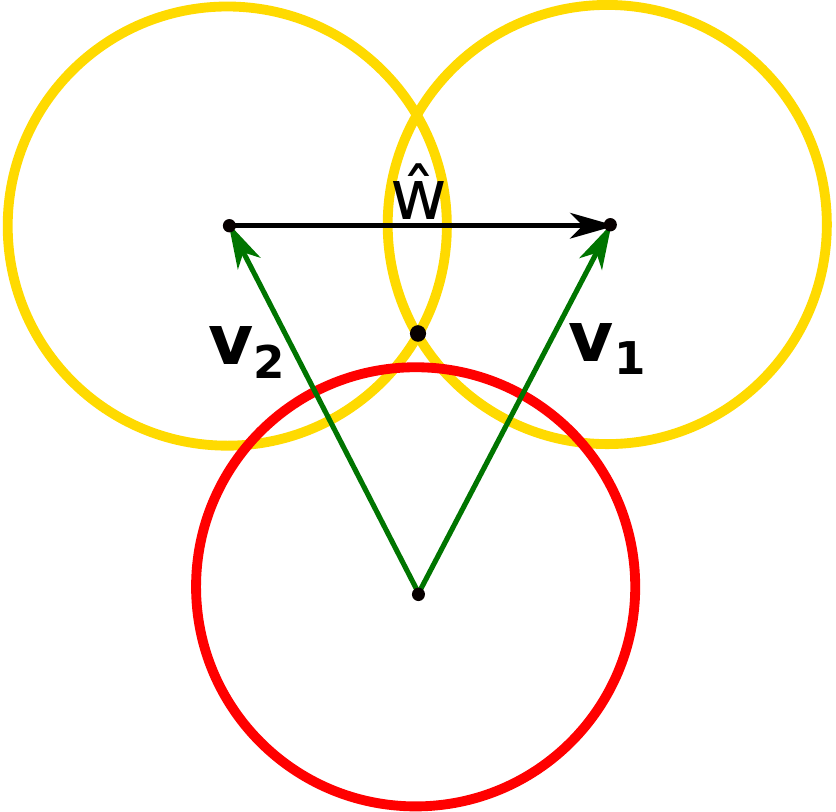}	
		\caption{\label{Fig:DCcontactdetection2} Two vectors $\mathbf{v_1}$ and $\mathbf{v_2}$ from the contacting red sphere's centre to the centres of the constituting spheres of the yellow dimer are determined. The projections of these two vectors onto the unit normal $\mathbf{w}$ of the circle enclosing cusp are determined and the directions of these projections are checked. (a) If both of them have the same direction, it is identified as a double contact (b) otherwise it is regarded as Type 2 configuration.}
	\end{figure}

	Double and cusp contacts are identified by checking if there is any overlap between the circle enclosing the cusp on the dimer surface and a contacting sphere of its neighbouring dimer, see Fig.~\ref{Fig:DCcontactdetection1}(a). This circle with centre $\mathbf{c_c}$, radius $r_c$ and unit normal $\mathbf{w}$ and a sphere with centre $\mathbf{c_s}$, radius $r_s$ are shown in Fig.~\ref{Fig:DCcontactdetection1}(b). The next steps are followed for the identification:
	
	\begin{enumerate}
		\item The distance $d_{cs}=|\mathbf{w}\cdot(\mathbf{c_c}-\mathbf{c_s})|$ between the plane of the circle and the sphere's centre is calculated to check if the plane cuts the sphere or not. 
		If $d_{cs}>r_s$ then there is no intersection, so the plane passes above/below the sphere entirely.
		\item  If there is an intersection, i.e., $d_{cs}<r_s$, it will be between the original circle and a new one formed where this plane meets the sphere, with centre $\mathbf{c_p}=\mathbf{c_s}+d_{cs}\mathbf{w}$.
		
		\item If $d_{cs}=r_s$ then this is the sole point of intersection with the plane, otherwise a new circle with radius $r_p$ occurs as displayed in Fig.~\ref{Fig:DCcontactdetection1}(c), where $r_p=\sqrt{{r_s}^2-d_{cs}^2}$. Then, the problem has been reduced to a circle-circle interaction. 
		
		\item If $|\mathbf{c_p}-\mathbf{c_c}|<r_c+r_p$, then there is overlap between the circle and the sphere, so the contact is identified as a cusp contact. If there is no overlap, then the contact is either a double contact or a Type 2 configuration. 
		
		\item To distinguish a double and a Type 2 configuration, two vectors $\mathbf{v_1}$ and $\mathbf{v_2}$ from the contacting sphere's centre to the centres of the constituting spheres of the reference dimer are determined as illustrated in Fig.~\ref{Fig:DCcontactdetection2}. The projections of these two vectors onto the unit normal $\mathbf{w}$ of the circle enclosing cusp are determined and the directions of these projections are checked. If both of them have the same direction, the contact is identified as a double contact, otherwise it is regarded as a Type 2 configuration.
	\end{enumerate}
	
	\begin{figure}[h]
		\centering
		\includegraphics[width=0.45\textwidth]{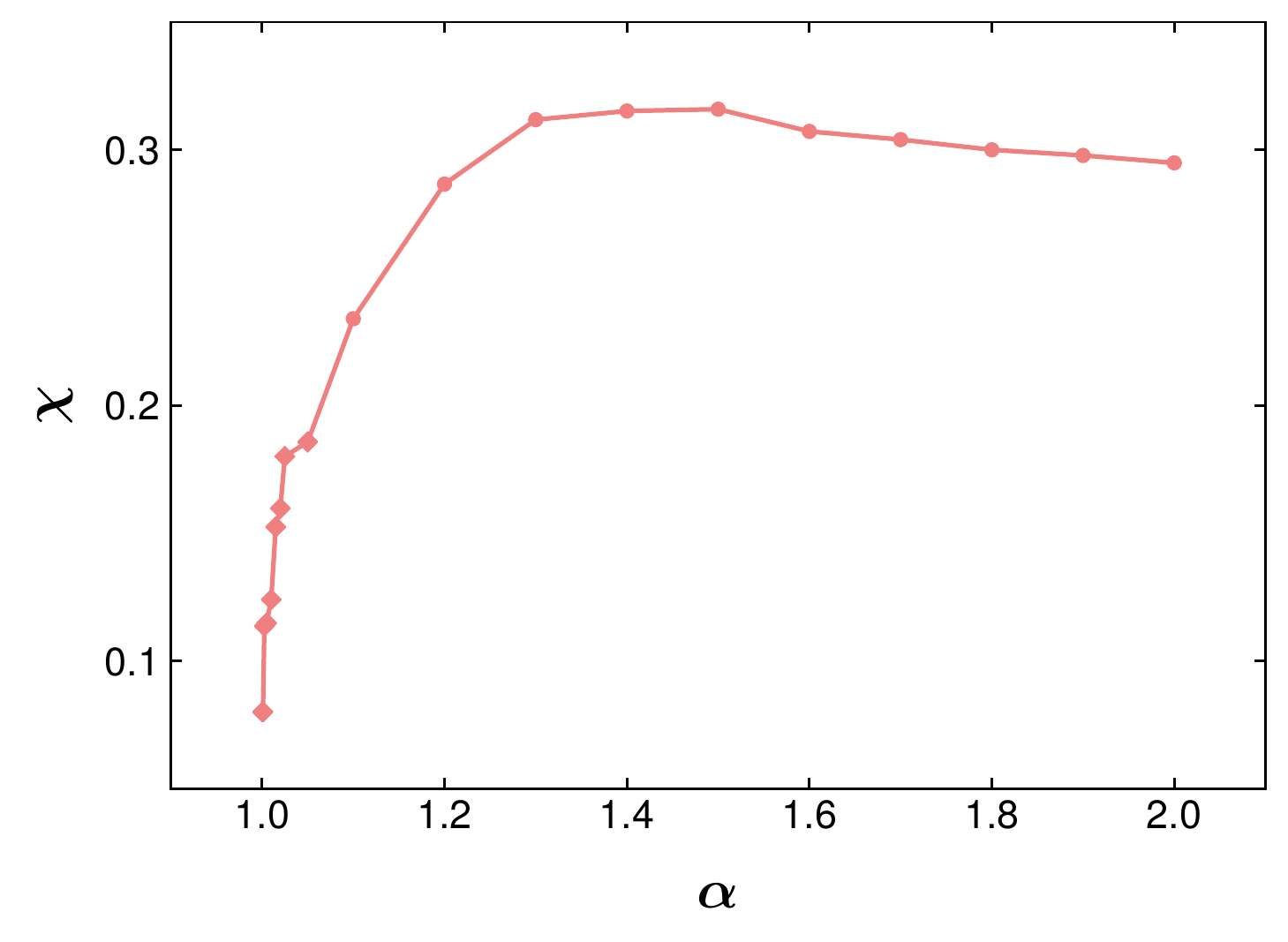}
		\caption{\label{Fig:chi}The orientational order parameter $\chi$ vs $\alpha$. Values of $\chi$ are shown averaged over 10 independent simulation runs for $\alpha\ge 1.1$ (dots), and for a single run for $\alpha<1.1$ (diamonds).}
	\end{figure}

	\section{The order parameter $\chi$}
	
	In \cite{Buchalter1994} the following order parameter has been introduced to measure the orientational order of prolate ellipsoids
	
	\begin{equation}
		\label{chi}
		\chi=\frac{3}{2} \left\{ \frac{1}{N_{\rm b}}  \sum_{i}^{N_{\rm b}}\cos^2\left( \beta_i-\frac{\pi}{2}\right)- \frac{1}{3} \right\}=\frac{1}{N_{\rm b}}  \sum_{i}^{N_{\rm b}}P_2\left( \beta_i-\frac{\pi}{2}\right)
	\end{equation}
	where $\beta_i$ is the angle between the semi-major (long) axis of particle $i$ and the $\hat{\mathbf{z}}$-axis (gravity direction). Since the director identified with the $Q$-tensor in Sec.~\ref{Sec:S} is also aligned with the $\hat{\mathbf{z}}$-axis, the expression for $\chi$ is the same as that for $S$, Eq.~\eqref{Sdef}, apart from the shift $-\pi/2$ in the argument of $P_2$. The parameter $\chi$ of Eq.~\eqref{chi} thus takes values in the interval $[-2,1]$: when all particles are randomly oriented, $\chi=0$, while if all particles' long axes are oriented in the horizontal plane normal to the gravity direction $\chi=1$. When the long axes of particles are oriented along the gravity direction we have $\chi=-2$. A plot of $\chi$ as a function of $\alpha$ for our dimer packing data is shown in Fig.~\ref{Fig:chi}.	
	
	\begin{table*}
		\centering
		\caption{Two-dimensional illustrations of configurations with double and cusp contacts. These configurations are re-assigned to Type 1, 2, and 4 as indicated in the table}
		\label{table3}
		\begin{tabular*}{\textwidth}{@{\extracolsep{\fill}}cccc}
			\hline
			Configuration type  & \multicolumn{3}{c}{Re-assigned configuration type} \\
			\hline
			& Type 1 & Type 2 & Type 4 \\
			\hline
			Type 2 &  	
			\begin{minipage}{.25\textwidth}
				\includegraphics[width=0.6\textwidth]{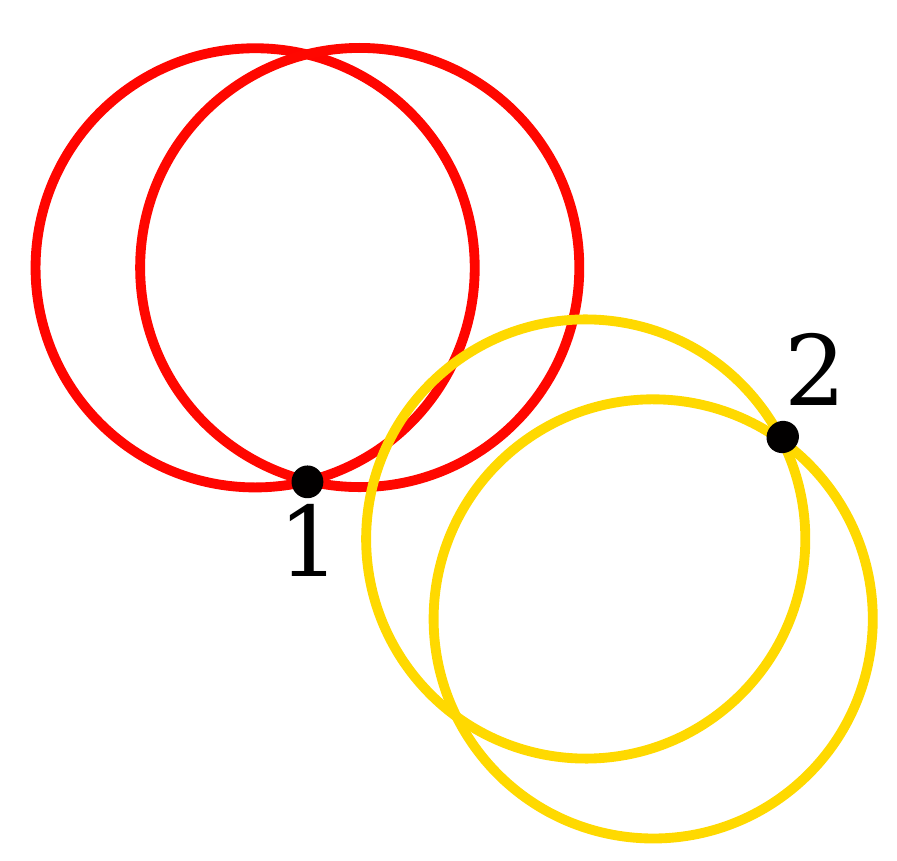}\\
				A double contact is counted as one contact point: two contact points are reduced to one.\vspace{0.2cm}
			\end{minipage} 	
			&  &   \\ 
			\hline
			Type 4 &
			\begin{minipage}{.25\textwidth}
				\includegraphics[width=0.6\textwidth]{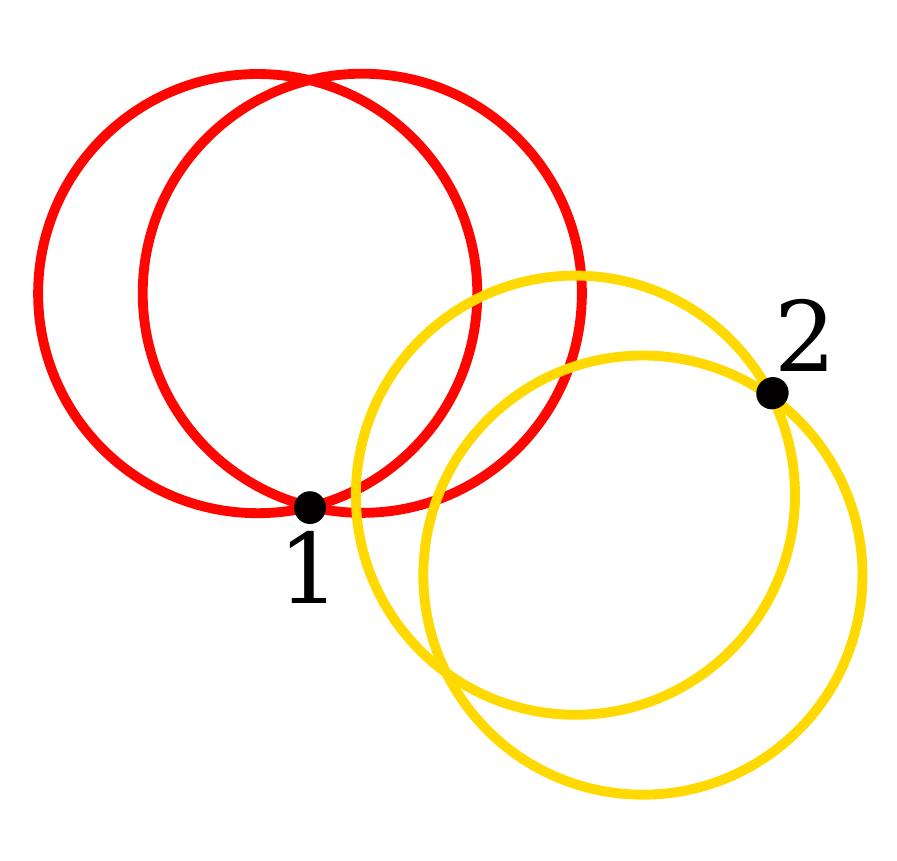}\\
				Two overlapping double contacts are counted as one contact point: three contact points are reduced to one.
			\end{minipage}
			&
			\begin{minipage}{.25\textwidth}
				\includegraphics[width=0.6\textwidth]{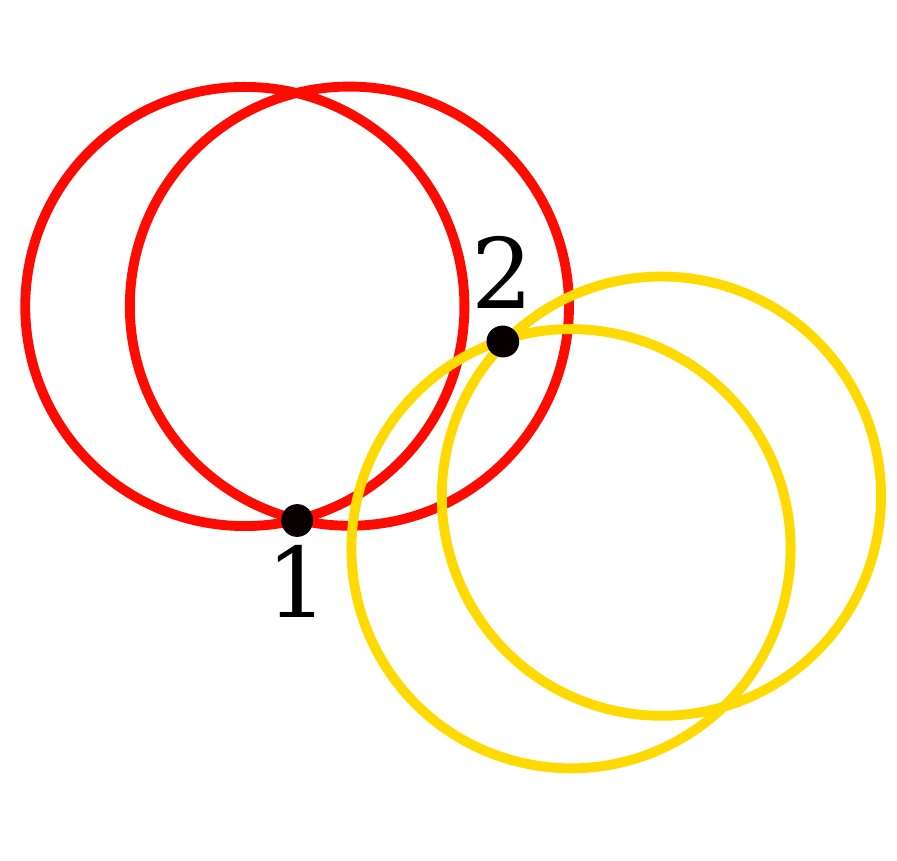}\\
				One double and one cusp contact (cusp 2 overlaps with the red sphere) are counted as two contact points: three contact points are reduced to two.\vspace{0.1cm}
			\end{minipage}
			&   \\ 
			\hline
			Type 5 &
			\begin{minipage}{.25\textwidth}
				\includegraphics[width=0.6\textwidth]{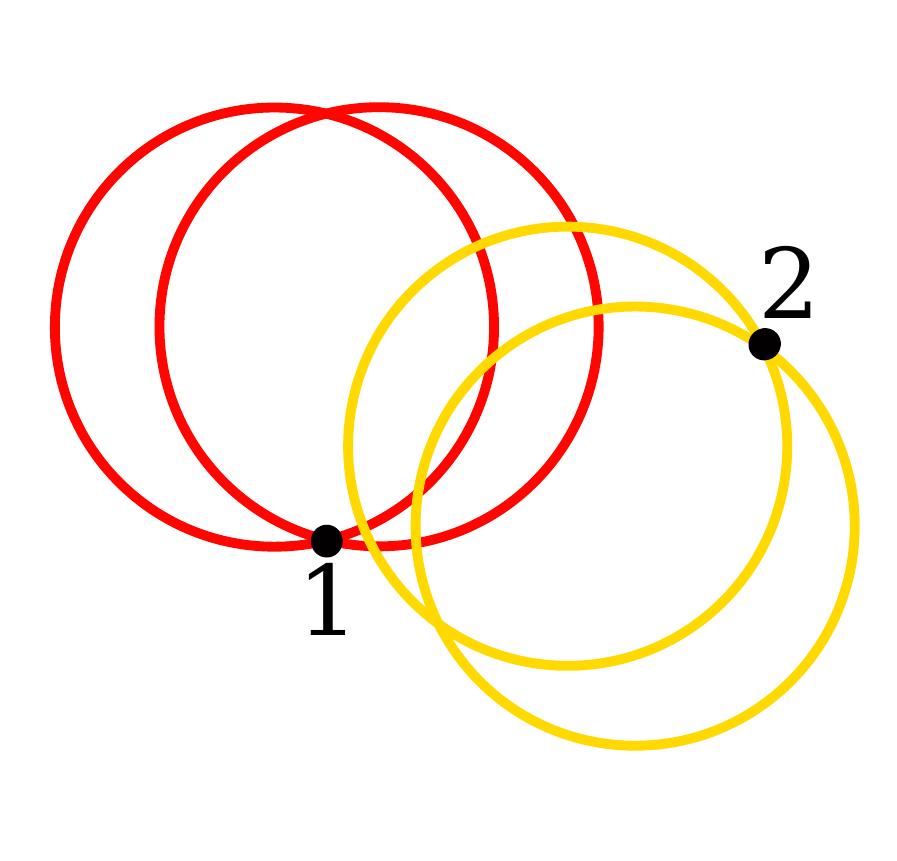}\\
				Two overlapping double contacts are counted as one contact point: four contact points are reduced to one.
			\end{minipage}		
			&
			\begin{minipage}{.25\textwidth}
				\includegraphics[width=0.6\textwidth]{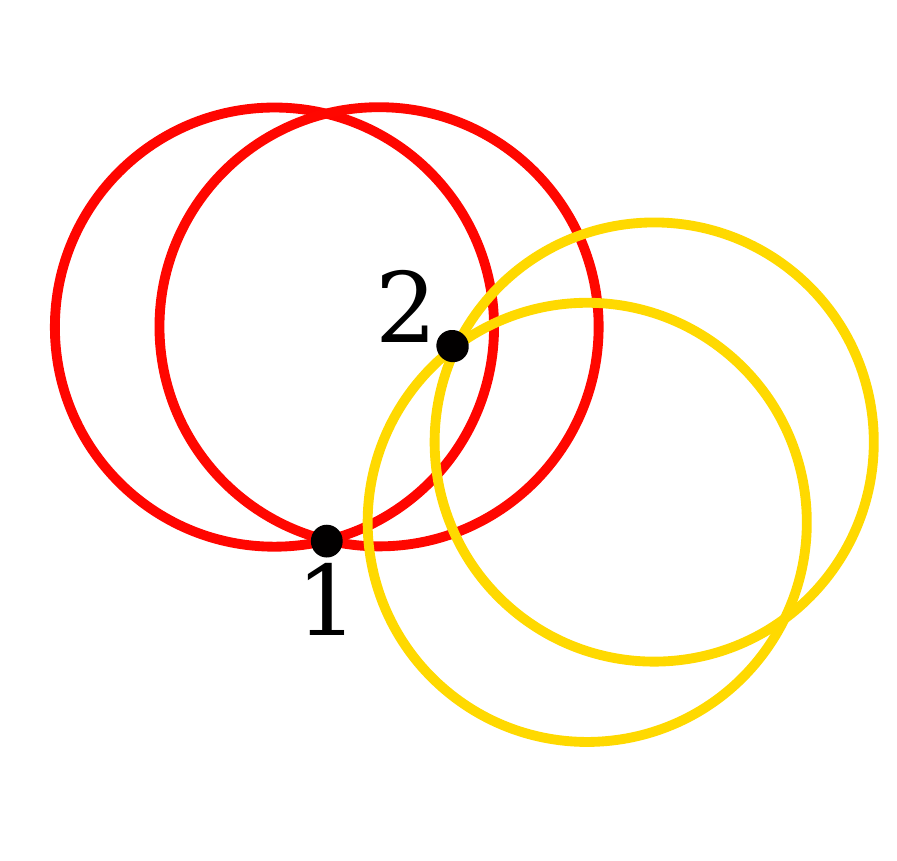}\\
				Two distinct double contacts are counted as two contact points: four contact points are reduced to two.
			\end{minipage}		
			&
			\begin{minipage}{.25\textwidth}
				\includegraphics[width=0.6\textwidth]{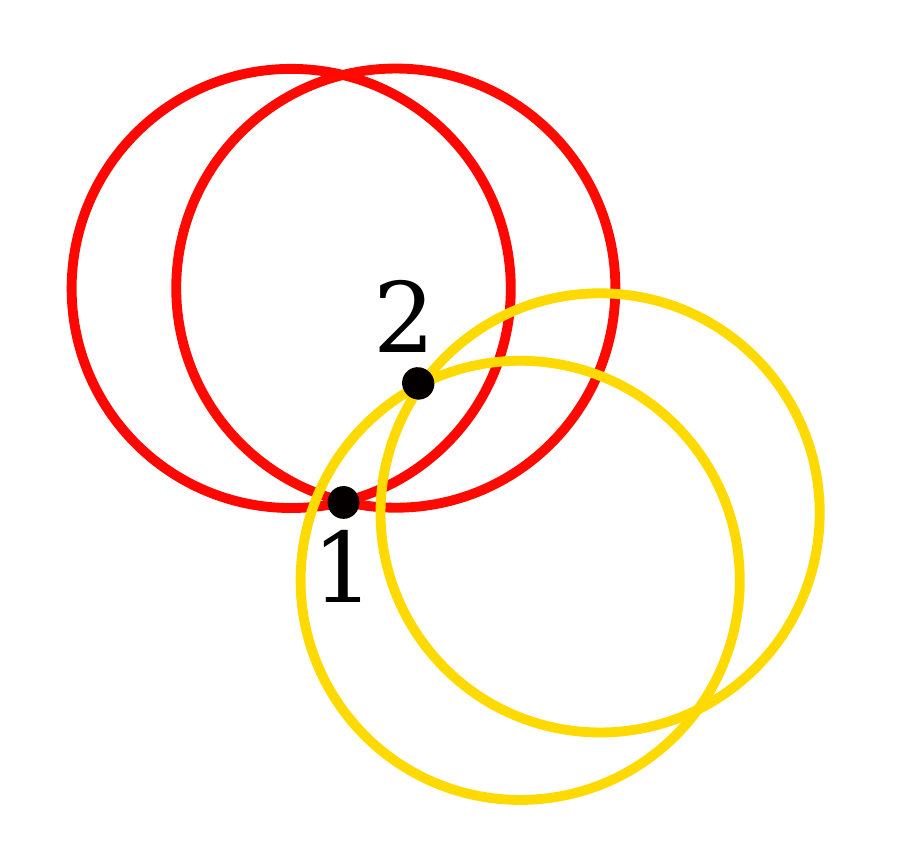}\\
				One double contact (cusp 1 is not covered by one of the yellow spheres) and one cusp contact (cusp 1 overlaps with the other yellow sphere) are counted as three contact points: four contact points are reduced to three.\vspace{0.1cm}
			\end{minipage}	    
			\\ 
			\hline		
		\end{tabular*}	
	\end{table*}
	
	\section{Mapping between different contact configuration types}
	\label{Sec:map}
	
	We introduce a heuristic method to re-assign configurations with double and cusp contacts to one of the Type 1, 2, and 4 configurations. The precise mapping depends on the number and the location of double and cusp contacts as summarized in Table~\ref{table3}. In general, double contacts are mapped to one contact point and cusp contacts to two. For Type 3 configurations, no double or cusp contacts have been found. For Type 5 configurations, two cusp contacts do occur, which leave the configuration as Type 5 after the mapping.
	
\begin{figure}[h]
		\centering
		\includegraphics[width=0.45\textwidth]{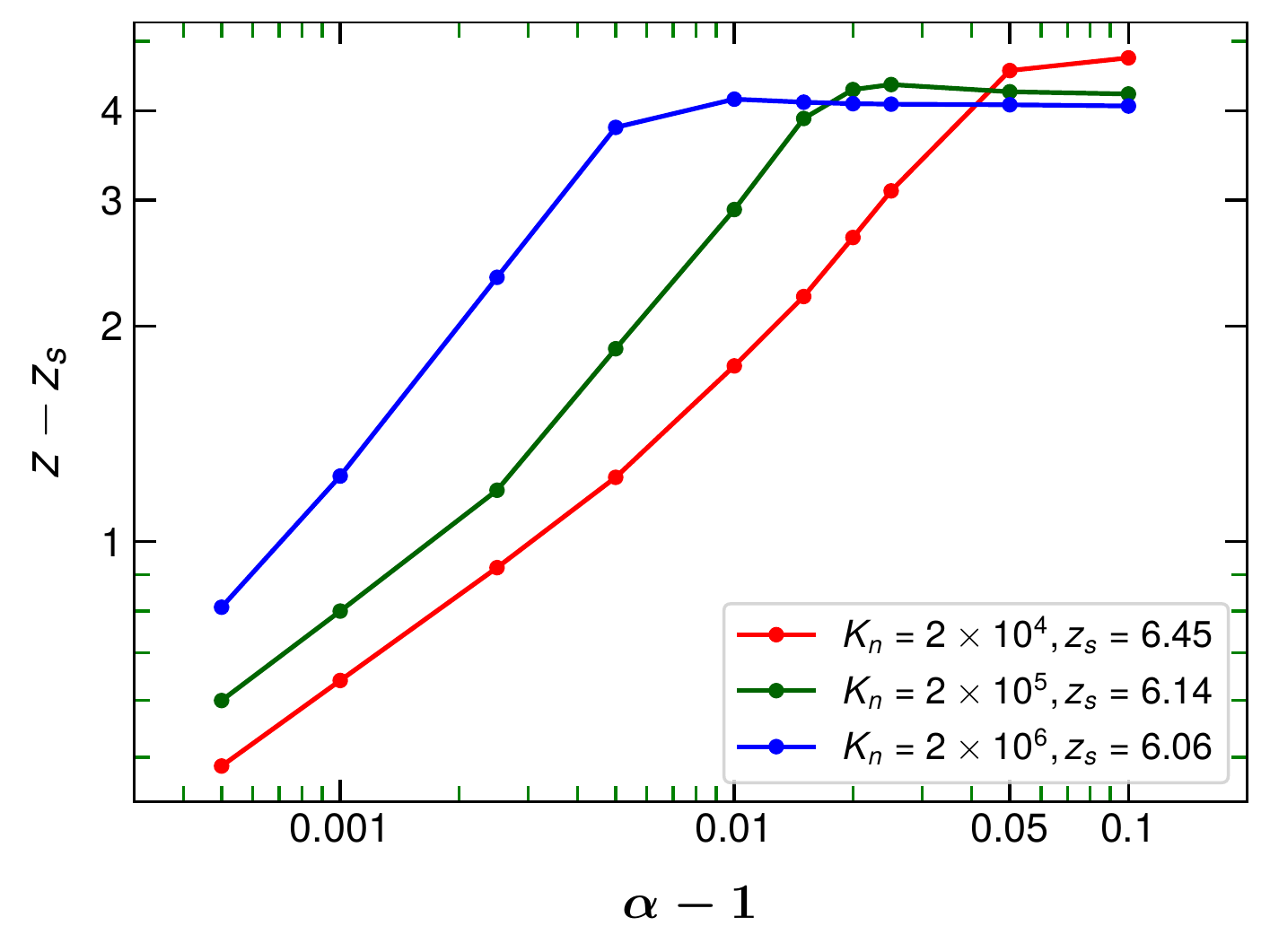}
		\caption{\label{Aspectratiocontactsmall}A double-logarithmic plot of $z-z_{\rm s}$ vs $\alpha-1$ for three different normal spring constants $K_{\rm n}$. We define $z_{\rm s}$ as the contact number of the corresponding sphere packing, which approaches the isostatic value $z_{\rm s}=6$ as the particle hardness increases. }
	\end{figure}

\begin{figure}[h]
		\centering 	
		\includegraphics[width=0.235\textwidth]{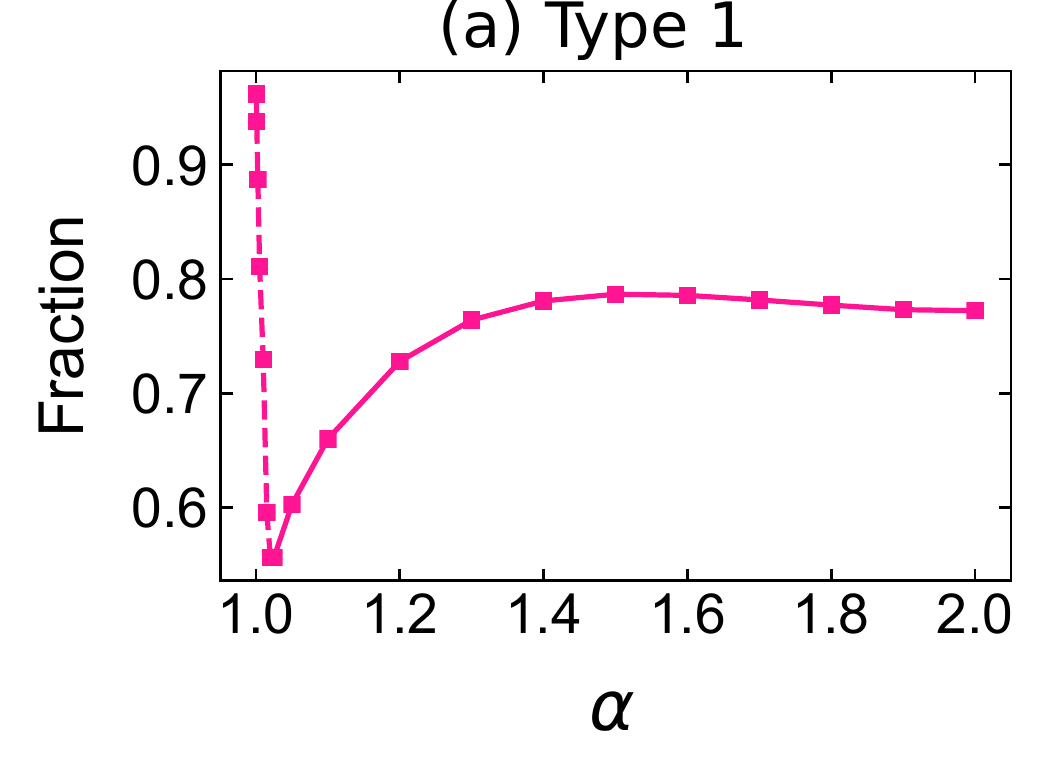} 
		\includegraphics[width=0.235\textwidth]{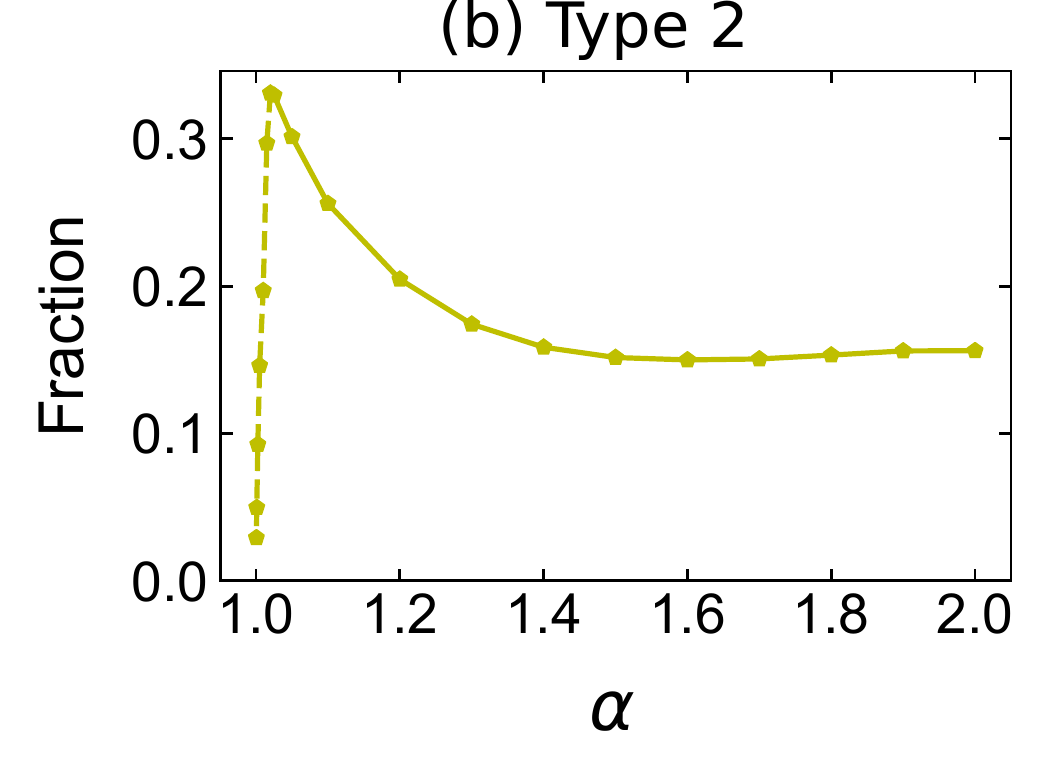} 
		\includegraphics[width=0.235\textwidth]{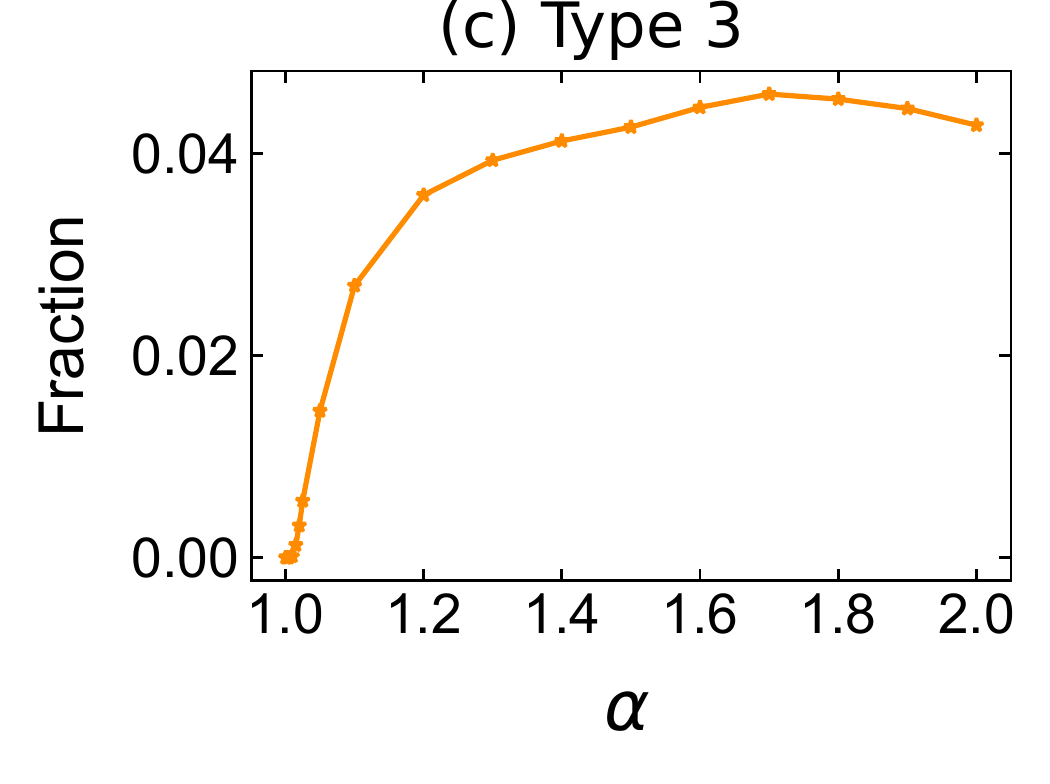}
		\includegraphics[width=0.235\textwidth]{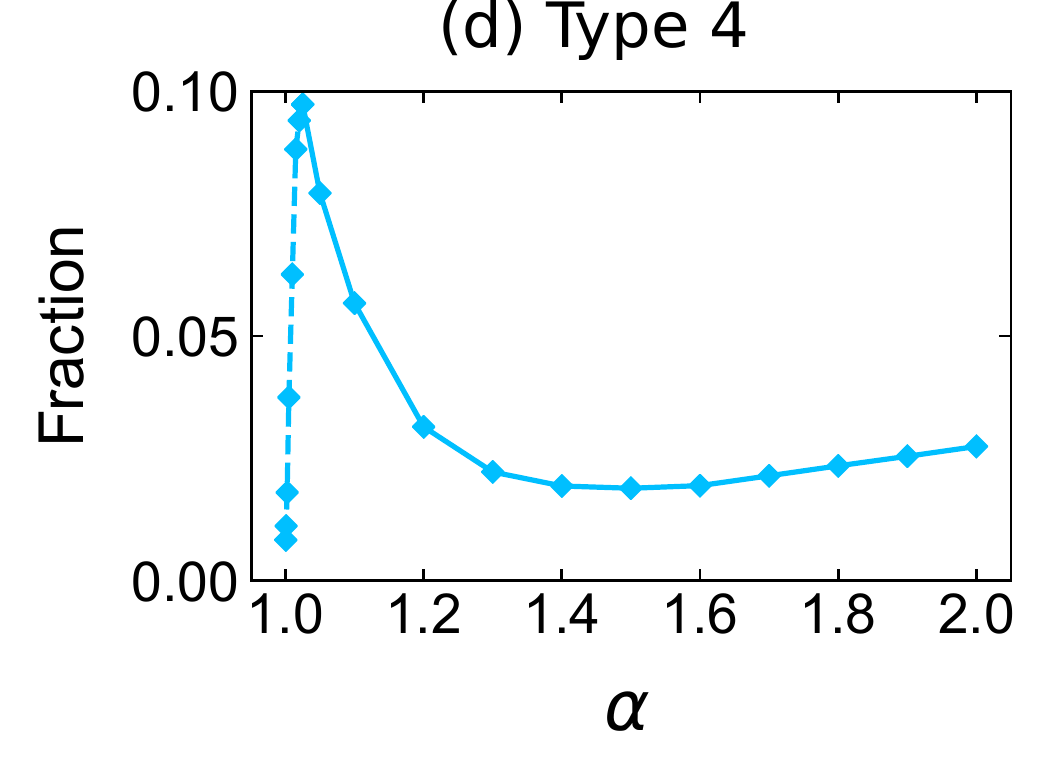} 
		\includegraphics[width=0.235\textwidth]{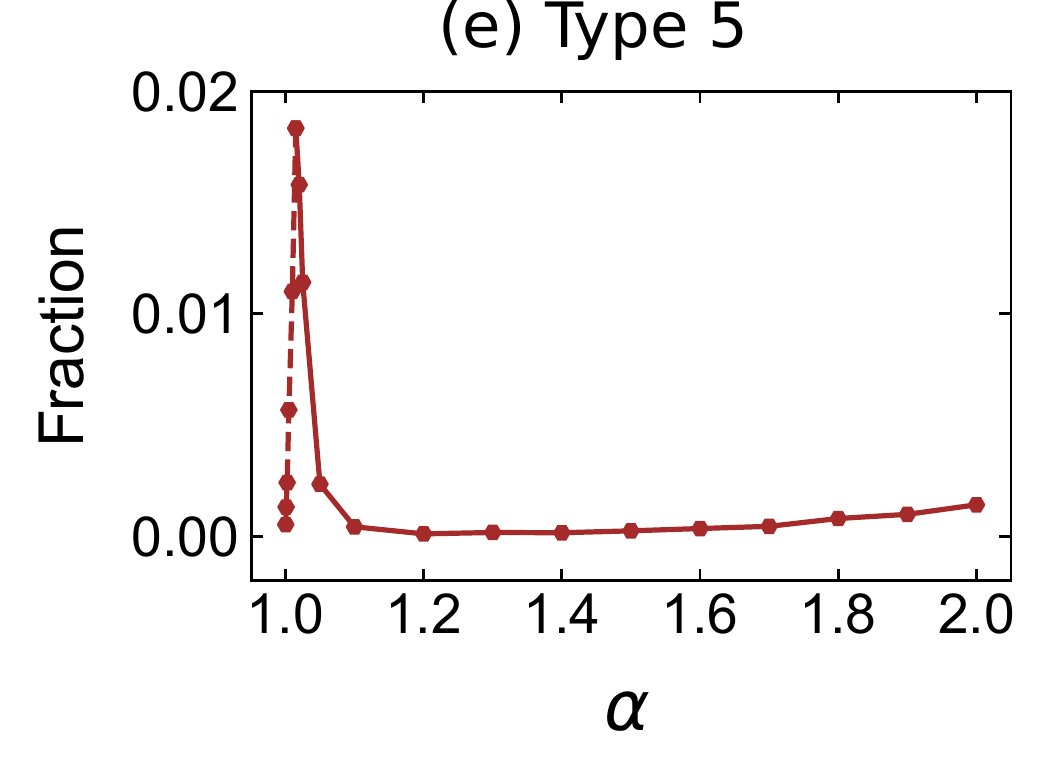}
		\caption{\label{Fig:Contacttypessmall}The fractions of the contact configurations of Type 1--5 vs. the aspect ratio $\alpha$. For $\alpha\ge 1.05$ the data shown is the same as in Fig.~\ref{Fig:Contacttypefraction}, but in the regime $\alpha< 1.05$ (dashed lines) the contact counting has been adjusted by re-assigning configurations with double and cusp contacts to Type 1, 2, and 4 configurations as summarized in Table~\ref{table3}.}
	\end{figure}

	With this mapping, we count a smaller number of contact points and thus the average number of contacts $z$ decreases. In fact, we obtain a rapid but smooth decrease of $z$ as $\alpha\to 1$, whereby $z$ approaches the corresponding value of spheres (Fig.~\ref{Aspectratiocontactsmall}). Resolving the contact counting by Type 1--5 configurations, we see that, as expected, the fraction of Type 1 configurations now increases for $\alpha<1.05$, while the fractions of Type 2,4, and 5 configurations decreases in the same regime (Fig.~\ref{Fig:Contacttypessmall}). In fact, the adjusted counting of contact points leads to sharp peaks at $\alpha\approx 1.05$, i.e., at the aspect ratio at which double and cusp contacts start to occur, that appear unphysical.

\end{appendix}

\end{document}